\documentclass{article}

\usepackage{arxiv}

\usepackage[utf8]{inputenc} 
\usepackage[T1]{fontenc}    
\usepackage{hyperref}       
\usepackage{url}            
\usepackage{booktabs}       
\usepackage{amsfonts}       
\usepackage{nicefrac}       
\usepackage{microtype}      
\usepackage{lipsum}
\usepackage{graphicx}

\usepackage[utf8]{inputenc}
\usepackage[OT1]{fontenc}
\usepackage{graphicx,xcolor}
\usepackage{amsmath}
\usepackage{amsfonts}
\usepackage{amssymb}
\usepackage{comment}
\usepackage{caption}
\usepackage{subcaption}
\usepackage{cancel}

\newcommand{\bm}{\textbf{\emph{m}}}
\newcommand{\brr}{\textbf{\emph{r}}}
\newcommand{\bs}{\textbf{\emph{s}}}
\newcommand{\bt}{\textbf{\emph{t}}}
\newcommand{\bu}{\textbf{\emph{u}}}
\newcommand{\bv}{\textbf{\emph{v}}}
\newcommand{\bx}{\textbf{\emph{x}}}
\newcommand{\bp}{\textbf{\emph{p}}}

\newcommand{\by}{\textbf{\emph{y}}}

\newcommand{\bl}{\boldsymbol{\ell}}
\newcommand{\btheta}{\boldsymbol{\theta}}
\newcommand{\bzeta}{\boldsymbol{\zeta}}
\newcommand{\bxi}{\boldsymbol{\xi}}
\newcommand{\bGamma}{\boldsymbol{\Gamma}}
\newcommand{\bPi}{\boldsymbol{\Pi}}

\newcommand{\bA}{\textbf{\emph{A}}}

\newcommand{\bC}{\textbf{\emph{C}}}
\newcommand{\bH}{\textbf{\emph{H}}}
\newcommand{\bI}{\textbf{\emph{I}}}

\newcommand{\bM}{\textbf{\emph{M}}}

\newcommand{\bQ}{\textbf{\emph{Q}}}

\newcommand{\bT}{\textbf{\emph{T}}}

\newcommand{\bW}{\textbf{\emph{W}}}
\newcommand{\bX}{\textbf{\emph{X}}}

\newcommand{\RR}{\mathbb{R}}
\newcommand{\PP}{\mathbb{P}}

\title{Crater Projection in Linear Pushbroom Camera Images\thanks{An earlier version of this manuscript was presented as Paper SIW22-33 at the \(3^{rd}\) Space Imaging Workshop, Atlanta, GA, 10-12 October 2022.}}

\author{
 \Large{Michela Mancini, Ava Thrasher, Carl De Vries and John Christian} \\
 \\
 Guggenheim School of Aerospace Engineering\\
  Georgia Institute of Technology\\
  Atlanta, GA, 30364
}

\begin{document}
\maketitle
\begin{abstract}
Scientific imaging of the Moon, Mars, and other celestial bodies is often accomplished with pushbroom cameras. Craters with elliptical rims are common objects of interest within the images produced by such sensors. This work provides a framework to analyze the appearance of crater rims in pushbroom images. With knowledge of only common ellipse parameters describing the crater rim, explicit formulations are developed and shown to be convenient for drawing the apparent crater in pushbroom images. Implicit forms are also developed and indicate the orbital conditions under which craters form conics in images. Several numerical examples are provided which demonstrate how different forms of crater rim projections can be interpreted and used in practice. 
\end{abstract}

\section{Introduction}\label{sec:introduction}

    Pushbroom cameras are common science instruments, with a long history of imaging the Moon, Mars, and other planetary bodies. Contemporary examples of pushbroom cameras on planetary exploration missions include the Mars Express High Resolution Stereo Camera (HRSC) \cite{Neukum:2004}, the Mars Reconnaissance Orbiter (MRO) High Resolution Imaging Science Experiment (HiRISE) \cite{McEwen:2007}, and the Lunar Reconnaissance Orbiter Camera's (LROC) Narrow Angle Camera (NAC) \cite{Robinson:2010}. Pushbroom images are formed as a sequence of projective one-dimensional (1-D) images captured by a linear array of detectors. The two-dimensional (2-D) image is formed as the camera passes over the terrain. The pushbroom camera is aptly named for the sensor's motion as it sweeps over the area being imaged. 
    
    Pushbroom cameras are not practical for real-time applications (e.g., optical navigation) because 2-D images are assembled as a sequence of 1-D scans over an extended time period. They do, however, provide several benefits for capturing scientific images for post-flight analysis. Historically, pushbroom cameras had the advantage of not containing moving mechanical parts which may fail during the mission lifetime. However, this oft-cited advantage is less important today with the widespread incorporation of digital shutter technologies in conventional cameras. Instead, the most important modern advantage is that pushbroom cameras tend to collect more radiometrically accurate images.
    Each detector in the linear array is constantly exposed to the scene (compared to framing cameras which have finite exposure times). These longer exposure times (or ``dwell times'') can result in a larger energy captured on individual detectors---which is particularly important under poor or variable illumination conditions \cite{Jensen:2007}. 
    In practice, the raw pushbroom images are downlinked to Earth and refined offline (e.g. with radiometric correction and applying standard map projections). The construction of world maps or digital elevation maps (DEMs) are two common data products produced from pushbroom images.
    
\begin{figure}[h]
    	\centering
    	\includegraphics[width=0.50\textwidth]{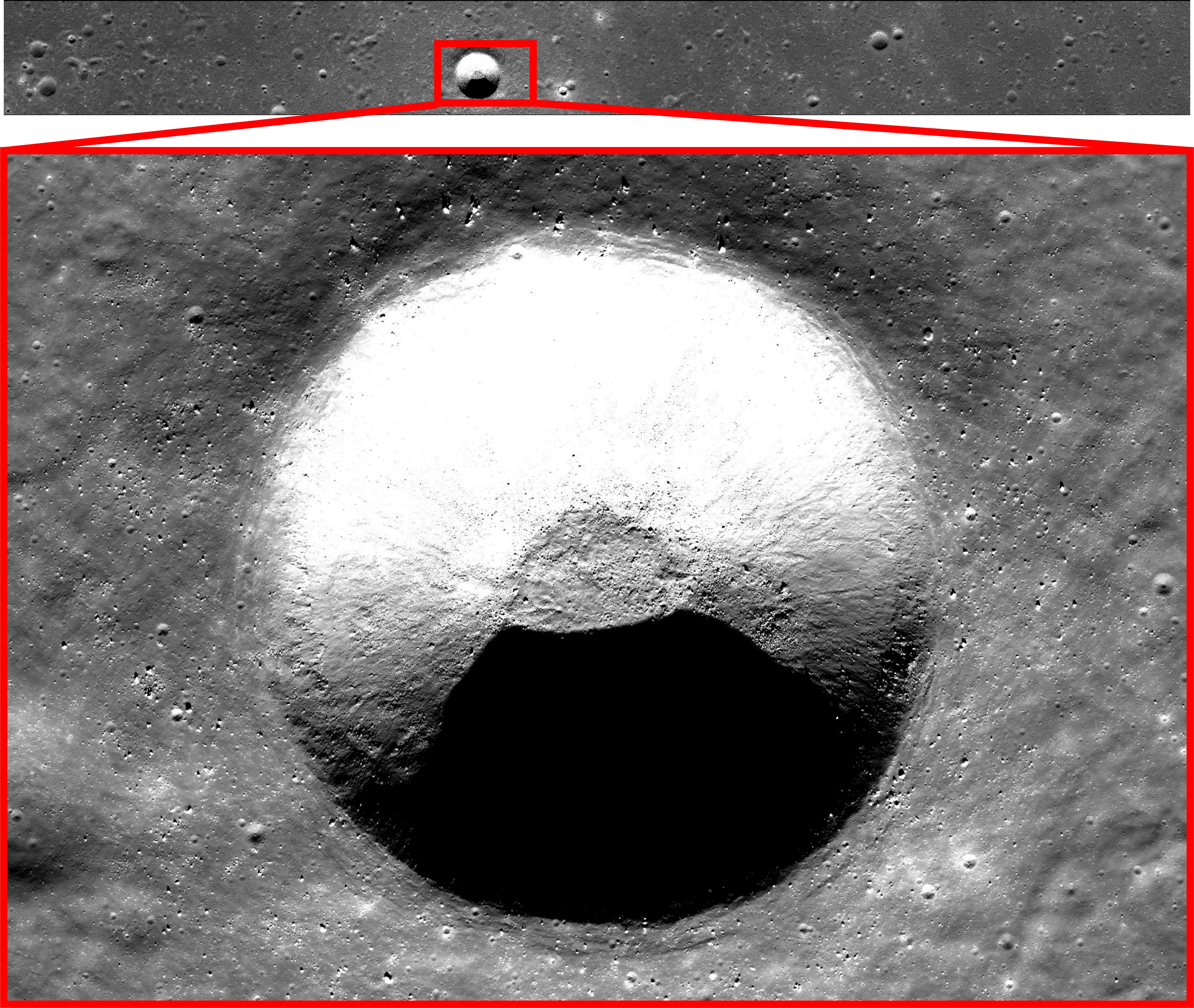}
    	\caption{Pushbroom cameras tend to produce high aspect ratio images with the along-track dimension much larger than the cross-track dimension due to the observer's orbital motion. For scale, Curtis crater is approximately 2.9 km in diameter. NASA PDS product \texttt{M1107903421RE} \cite{Robinson:2010b}.}
    	\label{fig:nac_pushbroom_image}
    \end{figure}

    Most pushbroom imagery is captured from a low emission angle. One example is shown in Fig.~\ref{fig:nac_pushbroom_image}, where we see the raw pushbroom image of a terrain patch in Mare Crisium across the top with a magnified view of Curtis crater below.  
    This is one example of the Moon's large population of craters. Many of these craters (as well as those on Mars) have been cataloged \cite{Robbins:2018, Lagain:2021}.
    While no crater is perfectly elliptical, analysis has shown that the assumption fits well with current observations \cite{Christian:2021, Bottke:2000}. 
    
    A planar conic (e.g., an ellipse) captured by a central perspective camera (modeled as a pinhole camera) is known to project to a conic in the image. The conic-to-conic mapping is governed by a homography which is well understood from projective geometry \cite{Hartley:2003}. In fact, this result has been utilized for several types of spacecraft optical navigation, such as crater-based \cite{Christian:2021} and horizon-based \cite{Christian:2021b} methods. Unfortunately, the analytical frameworks available for pushbroom cameras are not as mature as those for central projection cameras. Full pushbroom camera models are intractable for analytical analysis due to the complexities introduced by the spacecraft's orbital motion \cite{Gupta:1995}. The linear pushbroom model was developed to address some of these complexities \cite{Gupta:1997}, and it led to insights into the formation of pushbroom images, including the mapping of lines on a plane to hyperbolas in an image. 
    
    In this work, we describe an analytical framework to interpret the appearance of ellipses (e.g., craters) in pushbroom cameras. 
    We approach this work by first reviewing several mathematical formulations of ellipses. Specifically, explicit forms are developed for their convenient mapping from world coordinates to the pushbroom image plane as well as their practical use of drawing ellipses using only common ellipse parameters (e.g., semi-major axis and semi-minor axis). In addition, we utilize implicit forms to (1)~show that, in general, ellipses map to fourth-degree polynomials in pushbroom pixel coordinates and (2)~provide a comprehensive treatment of the observation geometry under which an ellipse maps to another conic in a pushbroom image. Finally, we show how the fourth-degree polynomial may be used to estimate the state (position and velocity) of the spacecraft, given the pushbroom image of a crater.
    
\section{Linear Pushbroom Cameras}\label{sec:linear_pushbroom_cameras}
    
\begin{figure}[b!]
    	\centering
    	\includegraphics[width = 0.5\textwidth]{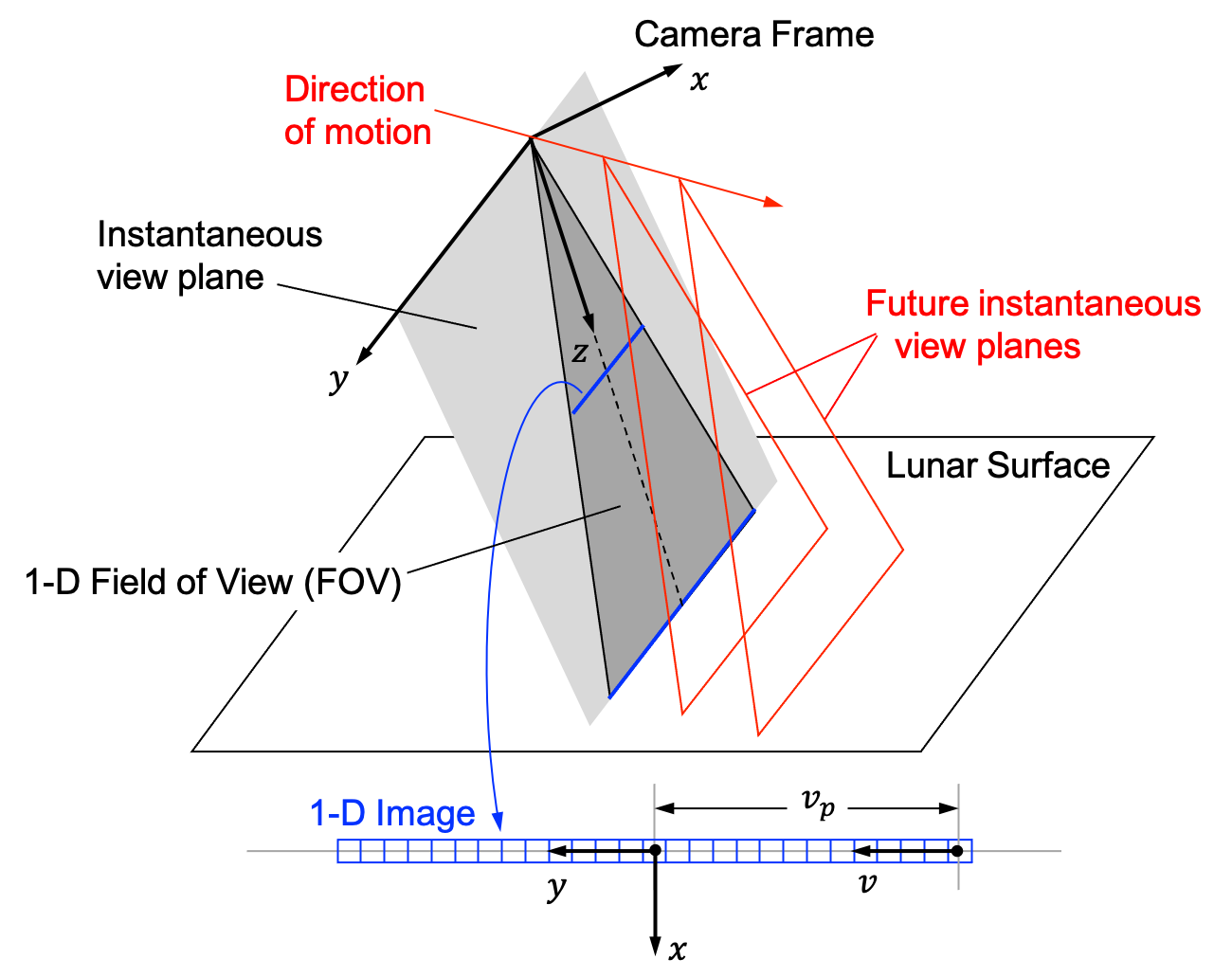}
    	\caption{The pushbroom camera's instantaneous view plane is the camera frame's $y$-$z$ plane and contains the 1-D image formed at any instant in time.}
    	\label{fig:PushbroomDef}
\end{figure}

A pushbroom camera is an optical sensor with a 1-D detector array following perspective projection. Thus, at any instant in time, a 1-D image is formed by the perspective projection of objects residing within the instantaneous view plane (see Fig.~\ref{fig:PushbroomDef}). The instantaneous view plane is formed by the join of the camera location (a point) and the 1-D  sensor strip (a line). As the sensor moves, the instantaneous view plane also moves and sweeps out a region of space. Stacking a sequence of 1-D images one after another creates a 2-D image.

Following the conventions of Hartley and Gupta~\cite{Gupta:1997}, we define a \emph{linear} pushbroom camera to describe the scenario when the camera moves at a constant velocity and maintains a constant attitude. We will now develop the linear pushbroom model.

Begin by defining the vector $\bl$ as the location of a world point $\bp$ relative to the camera at $\brr$,
\begin{equation}
    \label{eq:DefOfRelPosGeneric}
    \bl =  \bp - \brr 
\end{equation}
We choose to express everything in the known (and constant) basis vectors of the camera frame. This frame has the $z$-direction along the boresight direction and the $y$-direction along the 1-D sensor strip. Thus, the instantaneous view plane is the same as the camera frame $y$-$z$ plane (see Fig.~\ref{fig:PushbroomDef}).

Now, continue by considering the 1-D sensor at an instant in time. Since the 1-D sensor follows perspective projection, we find that
\begin{equation}
    \label{eq:1DPinhole}
    \begin{bmatrix}
        y \\ 1
    \end{bmatrix} \propto 
    \begin{bmatrix}
        0 & 1 & 0 \\ 0 & 0 & 1
    \end{bmatrix} \bl
\end{equation}
where $y$ is the coordinate where the direction $\bl$ pierces the $z=1$ plane. We observe that $[y, 1]^T \in \PP^1$.

We must now relate the point $y$ to its corresponding pixel location along the 1-D image. Define the $v$-direction to have units of pixels and to be along the 1-D image (and therefore parallel to the camera frame $y$-direction) as shown in Fig.~\ref{fig:PushbroomDef}. We then have a similarity transformation (corresponding to part of the usual 2-D pinhole camera calibration matrix \cite{Christian:2021b}) to transform from image plane coordinates to pixel coordinates
\begin{equation}
    \label{eq:1DCameraCalibrationTransform}
    \begin{bmatrix}
        v \\ 1
    \end{bmatrix} = 
    \begin{bmatrix}
        d_y & v_p \\ 0 & 1
    \end{bmatrix} 
    \begin{bmatrix}
        y \\ 1
    \end{bmatrix} 
\end{equation}
Hence, substituting Eq.~\eqref{eq:1DPinhole} into Eq.~\eqref{eq:1DCameraCalibrationTransform},
\begin{equation}
    \begin{bmatrix}
        v \\ 1
    \end{bmatrix} 
    \propto
    \begin{bmatrix}
        d_y & v_p \\ 0 & 1
    \end{bmatrix} 
    \begin{bmatrix}
        0 & 1 & 0 \\ 0 & 0 & 1
    \end{bmatrix} \bl
\end{equation}
\begin{equation}
    \begin{bmatrix}
        v \\ 1
    \end{bmatrix} 
    \propto
    \begin{bmatrix}
        0 & d_y & v_p \\ 0 & 0 & 1
    \end{bmatrix} 
    \bl
\end{equation}
We can remove the proportionality relation by introducing an explicit scaling, $w$,
\begin{equation}
    \label{eq:PerspectiveAxisModel}
    \begin{bmatrix}
        wv \\ w
    \end{bmatrix} 
    =
    \begin{bmatrix}
        0 & d_y & v_p \\ 0 & 0 & 1
    \end{bmatrix} 
    \bl
\end{equation}
which happens to be $w=\ell_z$.

The unusual part of the pushbroom camera model is how to handle the second dimension ($u$-direction) of the 2-D pushbroom image that is formed by the motion of the camera relative to the observed scene. Suppose we collect a sequence of 1-D images with a time $\tau$ between each image. Moreover, assume that each of these 1-D images forms a column in the resulting 2-D pushbroom image. In that case, the conversion from time to pixels is given by
\begin{equation}
    \label{eq:DefPushbroomu}
    u = (t - t_0) / \tau = \Delta t / \tau
\end{equation}
where \(t_0\) is the time at the beginning of the image, and perfect timing produces integer values of $u$.

A world point only appears in a particular 1-D image if it lies within the instantaneous view plane. Thus, if $\bv$ is the constant velocity of the observer, then the observer's instantaneous location at the time of a particular 1-D image is
\begin{equation}\label{eq:position_constant_velocity_kinematics}
    \brr = \brr_0 + \Delta t \, \bv
\end{equation}
where $\brr_0$ is the observer location at time $t_0$.
Substituting this result into Eq.~\eqref{eq:DefOfRelPosGeneric}, 
\begin{equation}
    \bl = \bp - \brr = \bp - (\brr_0 + \Delta t \, \bv) = \bl_0 - \Delta t \, \bv
\end{equation}
where $\bl_0 = \bp - \brr_0$. Writing out the terms and constraining $\bl$ to the instantaneous view plane (i.e., the camera $y$-$z$ plane, where $\ell_x=0$),
\begin{equation}
    \label{eq:LOSinInstViewPlane}
    \bl = \begin{bmatrix}
        0 \\ \ell_y \\ \ell_z
    \end{bmatrix} = 
    \begin{bmatrix}
        \ell_{0_x} \\ \ell_{0_y} \\ \ell_{0_z}
    \end{bmatrix} - \Delta t
    \begin{bmatrix}
        V_x \\ V_y \\ V_z
    \end{bmatrix}
\end{equation}
The first row gives us a relation between time, position, and velocity in the $x$-direction that allows us to determine when a point will pass through the instantaneous view plane
\begin{equation}
    \label{eq:Deltat1}
    \ell_{0_x} - \Delta t \, V_x = 0 \quad \rightarrow \quad
    \Delta t = \ell_{0_x} / V_x
\end{equation}
This can be transformed to the pushbroom image $u$ coordinate using Eq.~\eqref{eq:DefPushbroomu}
\begin{equation}
    u = \Delta t / \tau = \ell_{0_x} / (\tau V_x)
\end{equation}
or, in matrix form,
\begin{equation}
    \label{eq:Deltat2}
    u = \begin{bmatrix} 1 / (\tau V_x) & 0 & 0 \end{bmatrix} \bl_0
\end{equation}

To complete the camera model, we may use a similar scheme to analyze the $v$-coordinate of a moving pushbroom camera. Substituting $\Delta t$ from Eq.~\eqref{eq:Deltat1} into Eq.~\eqref{eq:LOSinInstViewPlane},
\begin{align}
    \bl = \begin{bmatrix}
        0 \\ \ell_y \\ \ell_z
    \end{bmatrix} & = 
    \begin{bmatrix}
        \ell_{0_x} \\ \ell_{0_y} \\ \ell_{0_z}
    \end{bmatrix} - \frac{\ell_{0_x}}{V_x}
    \begin{bmatrix}
        V_x \\ V_y \\ V_z
    \end{bmatrix}  \nonumber \\
    & = 
    \begin{bmatrix}
        0 & 0 & 0 \\
        -V_y/V_x & 1 & 0 \\
        -V_z/V_x & 0 & 1
    \end{bmatrix}
    \begin{bmatrix}
        \ell_{0_x} \\ \ell_{0_y} \\ \ell_{0_z}
    \end{bmatrix}
\end{align}
and further substituting into Eq.~\eqref{eq:PerspectiveAxisModel},
\begin{align}
    \label{eq:PerspectiveAxisModelv2}
    \begin{bmatrix}
        wv \\ w
    \end{bmatrix} 
    =
    \begin{bmatrix}
        d_y & v_p \\ 0 & 1
    \end{bmatrix} 
    \begin{bmatrix}
        -V_y/V_x & 1 & 0 \\
        -V_z/V_x & 0 & 1
    \end{bmatrix}
    \bl_0
\end{align}
Therefore, stacking Eqs.~\eqref{eq:Deltat2} and~\eqref{eq:PerspectiveAxisModelv2} yields one of the key relationships for pushbroom cameras:
\begin{align}
    \label{eq:PushbroomModel0}
    \begin{bmatrix}
        u \\ wv \\ w
    \end{bmatrix} 
    =
    \begin{bmatrix}
       1/\tau & 0 & 0\\0 & d_y & v_p \\ 0 & 0 & 1
    \end{bmatrix} 
    \begin{bmatrix}
         1/V_x & 0 & 0 \\
        -V_y/V_x & 1 & 0 \\
        -V_z/V_x & 0 & 1
    \end{bmatrix}
    \bl_0
\end{align}

The pixel coordinates \((u,\,v)\) might be related to the image plane coordinates through
\begin{equation}\label{eq:uv1TOxy1}
    \begin{bmatrix}
        u\\
        v\\
        1
    \end{bmatrix} = \begin{bmatrix}
        1/\tau & 0 & 0\\
        0 & d_y & v_p\\
        0 & 0 & 1
    \end{bmatrix}\begin{bmatrix}
        x\\
        y\\
        1
    \end{bmatrix}
\end{equation}
While this might be useful for image analysis, one should keep in mind that because of the orthographic projection along the \(x\)-direction
\begin{equation}
    \begin{bmatrix}
        x\\
        y\\
        1
    \end{bmatrix}\cancel{\propto} \,\lambda \begin{bmatrix}
        1/V_x & 0 & 0\\
        -V_y/V_x & 1 & 0\\
        -V_z/V_x & 0 & 1
    \end{bmatrix}\bl_0 \qquad \text{if } \lambda \neq 1
\end{equation}
Coming back to Eq.~\eqref{eq:PushbroomModel0}, we observe that $\bl_0$ is the vector from the initial camera location $\brr_0$ to the observed world point $\bp$, all expressed in the camera frame. To write this explicitly in another frame (e.g., a Moon-fixed frame $M$ in which elliptical crater rims are known),
\begin{align}
    \bl_0 = \bT^M_C (\bp_M - \brr_0) = 
    \begin{bmatrix}
        \bT^M_C & -\bT^M_C \brr_0
    \end{bmatrix} \bar{\bp}_M
\end{align}
where $\bp_M$ is the point expressed in the Moon-fixed frame and $\bar{\bp}_M^T = [\bp_M^T, 1]$. The $3 \times 3$ matrix $\bT^M_C$ is the attitude transformation matrix from the Moon-fixed frame $M$ to the pushbroom camera frame $C$. The relative position and orientation (sometimes called pose) of the camera relative to the Moon at time $t_0$ is given by the $3 \times 4$ frame transformation matrix   
\begin{align}
    \bPi^M_C = 
    \begin{bmatrix}
        \bT^M_C & -\bT^M_C \brr_0
    \end{bmatrix}
\end{align}
such that
\begin{align}
    \bl_0 = \bPi^M_C \bar{\bp}_M
\end{align}
Therefore, substituting into Eq.~\eqref{eq:PushbroomModel0},
\begin{align}
    \label{eq:PushbroomModel1}
    \begin{bmatrix}
        u \\ wv \\ w
    \end{bmatrix} 
    =
    \begin{bmatrix}
       1/\tau & 0 & 0\\0 & d_y & v_p \\ 0 & 0 & 1
    \end{bmatrix} 
    \begin{bmatrix}
        1 / V_x & 0 & 0 \\
        -V_y/V_x & 1 & 0 \\
        -V_z/V_x & 0 & 1
    \end{bmatrix}
    \bPi^M_C \bar{\bp}_M
\end{align}

\begin{figure}[hb!]
    	\centering
    	\includegraphics[width = 0.8\textwidth]{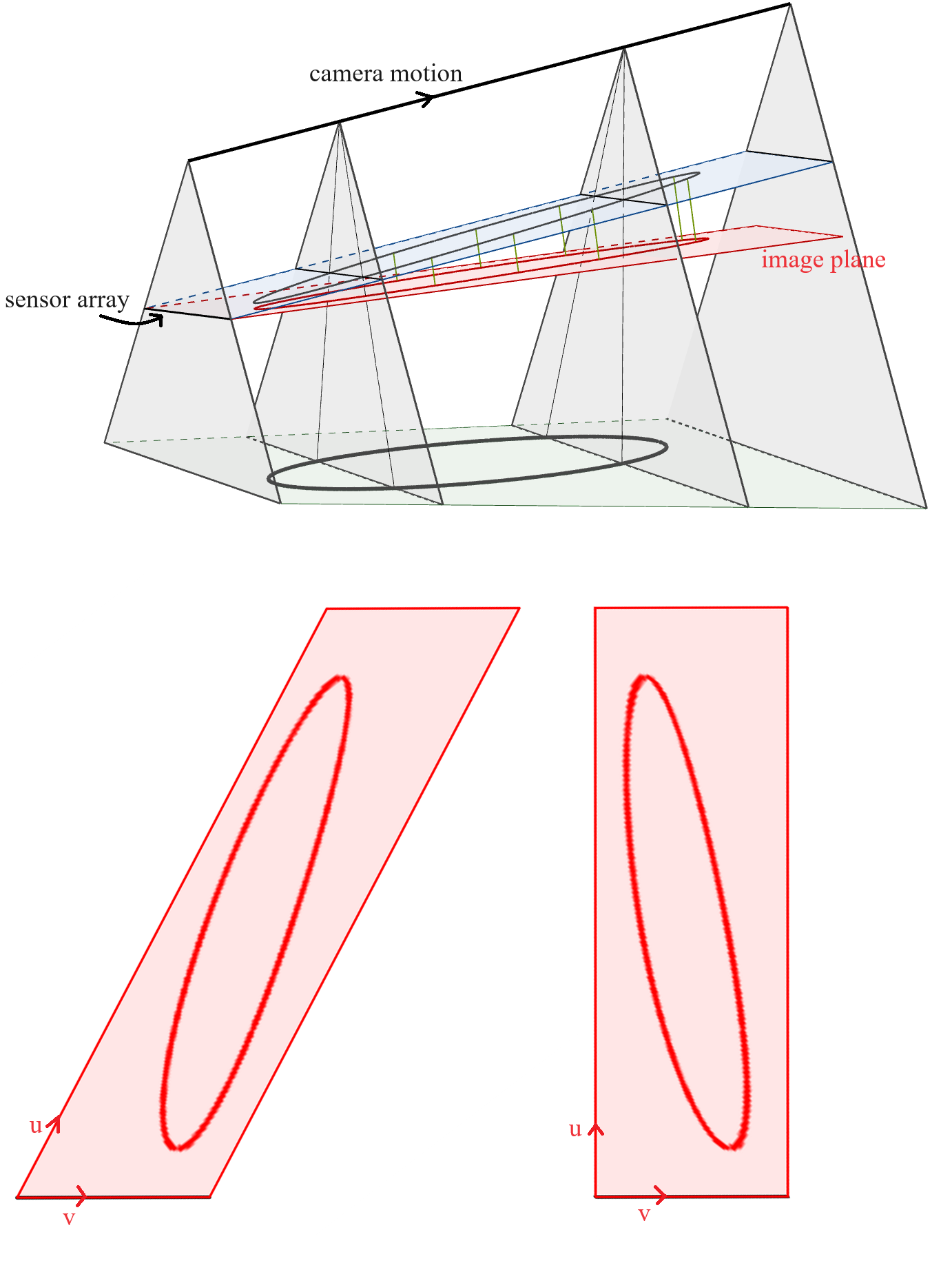}
    	\caption{The geometry of the projection on the image plane causes phenomena of shear and magnification that contribute to the appearance of the final pushbroom image. While the intersection between the lines of sight and the sensor array happens on the blue plane, the image plane is the red plane. In fact, while the blue plane has a constant distance from the camera center, only the red plane is orthogonal to the initial view plane.}
    	\label{fig:PushbroomDistortions}
\end{figure}

When the velocity is not purely along the camera frame $x$-direction and parallel to the ground, the camera motion will distort the resulting 2-D pushbroom image. In the simplest case, a vertical velocity (non-zero $V_z$) causes objects to appear larger or smaller and a horizontal velocity (non-zero $V_y$) causes image shear. This effect is illustrated in Fig.~\ref{fig:PushbroomDistortions}.

\section{Craters as Conics on the Lunar Surface}\label{sec:craters_as_conics_on_the_lunar_surface}
The pushbroom camera model developed in Eq.~\eqref{eq:PushbroomModel1} describes how a generic point $\bar{\bp}_M$ in the Moon frame (or, more generically, in any world frame) will project into an image. Suppose now that this point belongs to the rim of a crater. We want to develop a closed-form description of the curve that the projection of all the points on the crater rim forms in the image. Planetary crater rims are well-modeled as ellipses.  We will start by discussing the morphological features of common planetary surface craters and continue analyzing different conic parameterizations in order to identify the most convenient one, where by ``most convenient" we mean the one that allows us to easily develop the desired crater rim projection. 

\subsection{Craters formation and morphology}
Modeling crater rims as planar circles or ellipses is a well-established practice. There exist, in fact, multiple planetary crater catalogs where crater information is given in terms of radius and center coordinates (under the assumption of circular craters) or major and minor axes dimensions, center coordinates and crater orientation (under the assumption of elliptical craters). 

The shape of the crater is determined by multiple factors. As deeply studied in Ref.~\cite{Glass2013}, the formation of impact craters can be divided into three steps: contact and compression, excavation and modification. In all these phases, a pattern of compression and rarefaction waves, as well as the gravity of the planet, determine the final shape of the crater and its morphological features. When the impact angle is high (the trajectory of the impactor is not parallel to the surface of the planet), the wavefronts are approximately hemispherical, and the result is a circular crater rim \cite{Glass2013}. However, low impact angles produce different wavefronts, resulting in rims with higher ellipticity. Note that the threshold angle that determines when the ellipticity is high enough to make the crater classifiable as elliptical (usually a length-to-width ratio of at least 1.1) is highly material-dependent, as studied in \cite{Bottke:2000}. Here, the authors identify a threshold angle valid for Mars, Venus and the Moon, of \(12^{\circ}\). Despite this relatively low threshold, most lunar craters are nearly elliptical~\cite{Christian:2021}. Ref.~\cite{Robbins:2018} advanced multiple explanations for this, taking into account secondary impact craters and asymmetric erosion among other factors. For this reason, we decided to treat craters as elliptical objects, with dimensions and orientation provided by the Robbins crater database \cite{Robbins:2018}.

The energy of the impact drives the mechanism of crater formation, and may produce simple craters (less than 10-20 km), complex craters (more than 10-20 km) and basins (more than 300 km) \cite{Christian:2021}. Simple craters are usually bowl-shaped and have uplifted rims \cite{Christian:2021}. Complex craters are bigger, with a depth-to-diameter ratio that is smaller than that of simple craters. Some complex craters also exhibit one or more central peaks, due to the collapse of their walls in the modification phase \cite{Glass2013}. Others may have terraces or ring structures \cite{Christian:2021}.  Finally, craters of hundreds of kilometers are known as basins, and may have a complicated structure, with alternating concentric rings and valleys \cite{Glass2013}. As noticed in \cite{Christian:2021}, the morphological preservation of the crater will affect the quality of the ellipse fit.

\subsection{Conic parameterizations}
Since crater rims are well-modeled as ellipses~\cite{Christian:2021, Bottke:2000}, we can reasonably approximate observed points belonging to a crater rim as points constrained to lie on an ellipse---or, more generally, on a conic. The elliptical (conic) crater rims are assumed to be planar features.

Without loss of generality, we go back to our lunar example and define the Moon frame to have its origin at the center of the observed crater, with the $Z$-axis perpendicular to the plane of the crater and positive out of the surface (i.e., what someone sitting on the lunar surface would call the ``up'' direction). It follows that a point in the crater's plane is given by $\bp_M^T = [X,Y,0]$. Writing the 2-D location of a point on the elliptical crater rim in homogeneous coordinates as $\bar{\bs}^T = [X,Y,1] \in \PP^2$, we can write the following relation:
\begin{align}
    \label{eq:2DConicQuadricForm}
    \bar{\bs}^T \bC \bar{\bs} = 0
\end{align}
where $\bC$ is a symmetric matrix of ambiguous scale (five degrees-of-freedom) describing the conic locus \cite{Hartley:2003}. An equation of this form is true regardless of the orientation of the $X$-axis and $Y$-axis within the plane of the crater. However, it is often helpful for analytical analysis, to choose our basis vectors to simplify the problem. Thus, without loss of generality, place the origin at the ellipse center and define the $X$-axis to be in the direction of the elliptical crater's semi-major axis. In such a situation, points along the elliptical crater rim satisfy the constraint
\begin{equation}\label{eq:EllipsenInPlane}
    \frac{X^2}{a^2}+\frac{Y^2}{b^2}=1
\end{equation}
where $a$ is the semi-major axis and $b$ is the semi-minor axis. 
This parameterization is general enough to describe any ellipse and leads to a conic locus matrix of the form
\begin{equation}
    \label{eq:ConicLocusWorld}
    \bC\propto\begin{bmatrix}
            b^2 & 0 & 0\\
            0 & a^2 & 0 \\
            0 & 0 & -a^2b^2\\
        \end{bmatrix}
\end{equation}

The conic within the plane may be related to the 3-D geometry in a few different ways. To develop some useful relationships, begin by observing that
    \begin{equation}
    \label{eq:DefGamma}
        \bar{\bp}_M = 
        \begin{bmatrix}
            1 & 0 & 0 \\
            0 & 1 & 0 \\
            0 & 0 & 0 \\
            0 & 0 & 1
        \end{bmatrix} \bar{\bs} = \bGamma \bar{\bs}
    \end{equation}
and, since $\bGamma^T \bGamma = \bI_{3 \times 3}$, that
\begin{equation}
    \bar{\bs} = \bGamma^T \bar{\bp}_M 
\end{equation}
Substituting this result into Eq.~\eqref{eq:2DConicQuadricForm}, we find that
\begin{align}
    \bar{\bs}^T \bC \bar{\bs} = \bar{\bp}_M^T \bGamma \bC \bGamma^T \bar{\bp}_M = \bar{\bp}_M^T \bQ \bar{\bp}_M
\end{align}
where $\bQ$ is a $4 \times 4$ symmetric matrix of arbitrary scale that defines a quadric surface in $\PP^3$ \cite{Semple:1952}.
Taking the intersection of this quadric with the $z=0$ plane (i.e., the plane of the conic) leads to the conic section describing the crater rim.

The classical description of a planar conic in $\PP^2$ familiar to most analysts was given in Eq.~\eqref{eq:2DConicQuadricForm}.
    However, there exists an alternative called the canonical parametric representation which allows us to describe the 2-D coordinates of a conic using a single parameter \cite{Semple:1952}. This representation is sometimes more convenient than Eq.~\eqref{eq:2DConicQuadricForm}. 
    
    We can arrive at the desired single parameter representation using a simple transformation (a homography) on the projective plane $\PP^2$. 
    A homography is an isomorphism describing the mapping $\PP^n \rightarrow \PP^n$ that is usually parameterized by an $n \times n$ matrix of arbitrary scale. Consider a point $\bar{\bs} \propto [X,Y,1] \in \PP^2$ belonging to the elliptical crater rim and residing within the plane of the crater. Suppose there is a homography that relates $\bar{\bs}$ to a corresponding point $\bar{\btheta} \propto [\psi, \theta, 1] \in \PP^2$ in a transformed space. Thus, we may write 
    \begin{equation}
        \label{eq:ChangeOfCoordinates}
        \bar{\bs} \propto \bH \bar{\btheta}
    \end{equation}
    where $\bH$ is a $3 \times 3$ invertable matrix of arbitrary scale. 
    
    We know that a homography maps a conic to a conic \cite{Hartley:2003, Christian:2021}. If the goal is to reduce the conic description to a single parameter, a particularly convenient choice for the transformed conic is the parabola $\psi = \theta^2$. In this case, any point $\bar{\btheta}^T \propto [\psi, \theta, 1] = [\theta^2 \: \theta \: 1]$ lies on the parabola and depends only on the single parameter $\theta \in \RR$ \cite{Semple:1952}. This simple parabola can be written in terms of the usual quadric equation
     \begin{equation}
    \bar{\btheta}^T  \bC_{\theta} \bar{\btheta} = 2 \theta^2 - 2\psi = 0 
    \end{equation}
    where the conic locus matrix $\bC_{\theta}$ is given by
    \begin{equation}\label{eq:CTHETA}
        \bC_{\theta} \propto 
        \begin{bmatrix}
        0 & 0 & -1\\
        0 & 2 & 0\\
        -1 & 0 & 0\\
        \end{bmatrix}
    \end{equation}
    The objective, therefore, is to find the homography between the transformed conic locus $\bC_{\theta}$ and the original conic locus $\bC$.
    
    Substituting Eq.~\eqref{eq:ChangeOfCoordinates} into Eq.~\eqref{eq:2DConicQuadricForm} yields
    \begin{equation}
        \bar{\bs}^T \bC \bar{\bs} = \bar{\btheta}^T \bH^T \bC \bH \bar{\btheta} = \bar{\btheta}^T \bC_{\theta} \bar{\btheta} = 0 
    \end{equation}
    so that the transformation we seek is given by
    \begin{equation}\label{eq:ctheta}
        \bC_{\theta} \propto 
        \bH^T\,\bC\,\bH
    \end{equation}
    Given $\bC$ from Eq.~\eqref{eq:ConicLocusWorld} and $\bC_{\theta}$ from Eq.~\eqref{eq:CTHETA}, it is straightforward to show that the homography
    \begin{equation}
        \label{eq:HomographyForSingleParam}
        \bH \propto 
        \begin{bmatrix}
        a & 0 & -a\\
        0 & 2b & 0\\
        1 & 0 & 1\\
        \end{bmatrix}
    \end{equation}
    will produce the desired anti-diagonal matrix $\bC_{\theta}$. 
    
    Substituting the homography from Eq.~\eqref{eq:HomographyForSingleParam} into Eq.~\eqref{eq:ChangeOfCoordinates} yields
    \begin{equation}
        \bar{\bs} = \begin{bmatrix}
        X\\
        Y\\
        1
        \end{bmatrix} 
        \propto \begin{bmatrix}
        a(\theta^2-1)\\
        2b\theta\\
        \theta^2+1
        \end{bmatrix} = \bH \bar{\btheta}
    \end{equation}
    or, equivalently,
    \begin{equation}\label{eq:ParamXform}
        \bar{\bs} = \frac{1}{\theta^2+1} \bH \bar{\btheta}
    \end{equation}
    This allows us to explicitly write the ($X,Y$) coordinates belonging to the crater rim in terms of the parameter $\theta$  
    \begin{subequations}
    \begin{equation}\label{eq:XTheta}
        X=\frac{a(\theta^2-1)}{\theta^2+1}
    \end{equation}
    \begin{equation}\label{eq:YTheta}
        Y=\frac{2b\theta}{\theta^2+1}
    \end{equation}
    \end{subequations}
    
    A point around the ellipse may also be parameterized by the angle $\phi$
    \begin{equation}\label{eq:XYphi}
        X = a \cos \phi \quad \text{and} \quad Y = b \sin \phi
    \end{equation}
    which leads to
    \begin{equation}
        \label{eq:EllipseAngles}
        \cos \phi = \frac{(\theta^2-1)}{\theta^2+1} \quad \text{and} \quad \sin \phi =  \frac{2\theta}{\theta^2+1}
    \end{equation}
    Where we observe that
    \begin{equation}
        \cot \left(\frac{\phi}{2}\right) = \frac{1+\cos \phi}{\sin \phi} 
    \end{equation}
    and substitution from Eq.~\eqref{eq:EllipseAngles} shows that
    \begin{equation}\label{eq:thetaPhiRelation}
        \theta = \cot \left( \phi / 2 \right) 
    \end{equation}
    which provides a geometric understanding of how the parameter $\theta$ is related to the angle around the ellipse.
    
\begin{figure*}[h]
    	\centering
    	\includegraphics[width=\textwidth]{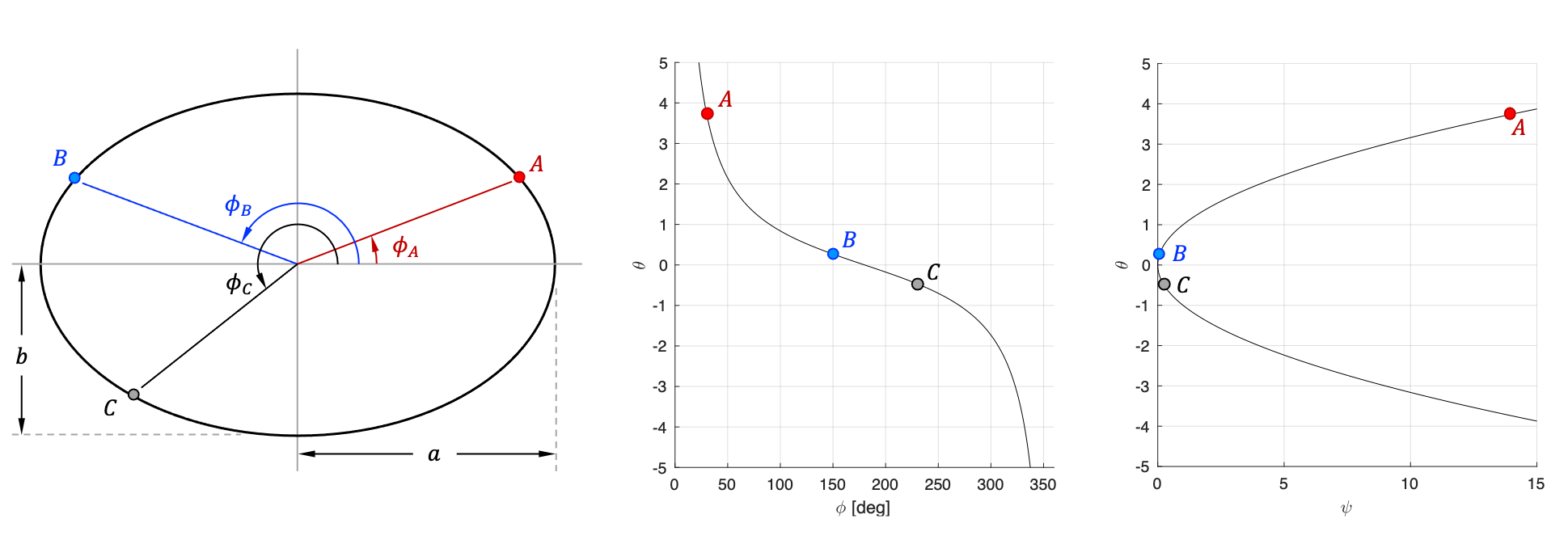}
    	\caption{The central angle which parameterizes the standard trigonometric form of an ellipse may be converted to the canonical representation parameter using a simple one-to-one cotangent mapping, and vice versa.}
    	\label{fig:EllipseParamExample}
    \end{figure*}

    To see these representations and transformations in action, consider an ellipse with $a=15$ and $b=10$ as shown in Fig.~\ref{fig:EllipseParamExample}. Choose three points $A,B,C$ around the ellipse occurring at angles $\phi_A = 30$ deg, $\phi_B = 150$ deg, and $\phi_C = 230$ deg. Substitution into Eq.~\eqref{eq:XYphi} gives the $(X,Y)$ coordinates of these three points, while substitution into Eq.~\eqref{eq:thetaPhiRelation} gives the parameter $\theta$ for the same three points. The reader may verify that substitution of the values $\theta_A,\theta_B,\theta_C$ into Eq.~\eqref{eq:XTheta} and ~\eqref{eq:YTheta}  recovers the corresponding ($X,Y$) coordinates. 

\section{Conics in Pushbroom Images} \label{sec:conics_in_pushbroom_images}
    
    Elliptical crater rims project to image conics with conventional perspective projection cameras (i.e., those that follow the pinhole camera model in both the $x$ and $y$ directions)\cite{Christian:2021}. However, elliptical crater rims do not necessarily project to conic features in a linear pushbroom image. To obtain a non-degenerate conic crater rim (e.g., ellipse) in a pushbroom image, the plane spanned by the camera $Y$-axis and the velocity vector must be parallel to the plane of the crater.
    All other motions or orientations lead to either a degenerate conic or a more complicated curve (generically a polynomial of degree four as discussed in subsequent sections). That we only obtain a non-degenerate image conic under the above conditions will now be shown.
    
    To begin, substitute Eq.~\eqref{eq:DefGamma} into Eq.~\eqref{eq:PushbroomModel1} to find
    \begin{equation}
        \label{eq:PushbroomMapping}
        \begin{bmatrix}
        u \\ wv \\ w
    \end{bmatrix}  = 
        \bM \bar{\bs}
    \end{equation}
    where $\bM$ is the $3 \times 3$ matrix
    \begin{align}
    \label{eq:DefM}
    \bM
    =
    \begin{bmatrix}
       1/\tau & 0 & 0\\0 & d_y & v_p \\ 0 & 0 & 1
    \end{bmatrix} 
    \begin{bmatrix}
        1 / V_x & 0 & 0 \\
        -V_y/V_x & 1 & 0 \\
        -V_z/V_x & 0 & 1
    \end{bmatrix}
    \bPi^M_C \bGamma
\end{align}
    Moreover, by introducing the diagonal matrix $\bW$,
    \begin{equation}
        \begin{bmatrix}
            u \\ wv \\ w
        \end{bmatrix}
        =
        \begin{bmatrix}
            1 & 0 & 0 \\
            0 & w & 0 \\
            0 & 0 & w
        \end{bmatrix}
        \begin{bmatrix}
            u \\ v \\ 1
        \end{bmatrix}
        =
        \bW \bar{\bu}
    \end{equation}
    we see that
    \begin{equation}
        \bar{\bs} = \bM^{-1} \bW \bar{\bu}
    \end{equation}
    Observe that $\bM$ is a constant matrix for any given pushbroom image, while $\bW$ may be different for every point (since $w$ may be different for every point). Proceeding undeterred, we find that
    \begin{align}
        \bar{\bs}^T \bC \bar{\bs} = \bar{\bu}^T \bW^T \bM^{-T} \bC \bM^{-1} \bW \bar{\bu} = 0
    \end{align}
    Writing this more compactly,
    \begin{align}
        \bar{\bu}^T \bA \bar{\bu} = 0
    \end{align}
    where
    \begin{align}\label{eq:ConicMatrixUVPlane}
        \bA = \bW^T \bM^{-T} \bC \bM^{-1} \bW
    \end{align}
    When the matrix $\bA$ is constant, the points $\bar{\bu}$ trace out a conic. Since $\bM$ and $\bC$ are constants, one obtains a conic when $\bW$ is constant---and this occurs when $w$ is constant. Consequently, we see that an elliptical crater will project to a conic in a pushbroom image when $w$ is constant. Recall from Eq.~\eqref{eq:PushbroomModel0}, that the scaling $w$ is given by
    \begin{equation}
        \label{eq:w_equation}
        w = \ell_{z} =  -\ell_{0_x}\frac{V_z}{V_x} + \ell_{0_z}
    \end{equation}
    
    The question at hand is: under what conditions is $w$ constant? The answer has two parts. The first part considers points on the conic observed at any one instant in time. The second part considers points observed at different times. 
    Begin by considering two points observed at the same time. When viewing a crater, the instantaneous view plane can intersect the conic in no more than two places. For the projection to be a conic in the pushbroom image, the value of $w$ for both of these points must be the same. If two conic points are from the same instant in time they have the same $u$ coordinate. Thus, from the first row of Eq.~\eqref{eq:PushbroomModel0},
    \begin{equation}
        u = \ell_{0_{x_1}}/(\tau V_x) = \ell_{0_{x_2}}/(\tau V_x) \; \rightarrow \; \ell_{0_{x_1}} = \ell_{0_{x_2}}
    \end{equation}
    And so it follows from Eq.~\eqref{eq:w_equation} that $\ell_{0_{z_1}} = \ell_{0_{z_2}}$. Only the $\ell_{0_y}$ coordinates of the two points may be different. Consequently, the line of intersection between the instantaneous view plane (i.e., camera $y$-$z$ plane) and the crater plane must be purely in the camera frame $y$-direction. This only happens when the $y$-direction is parallel to the plane of the crater (i.e. when the camera $x$-$z$ plane is perpendicular to the plane of the crater).
    
\begin{figure}[h]
    	\centering
    	\includegraphics[width=0.6\textwidth]{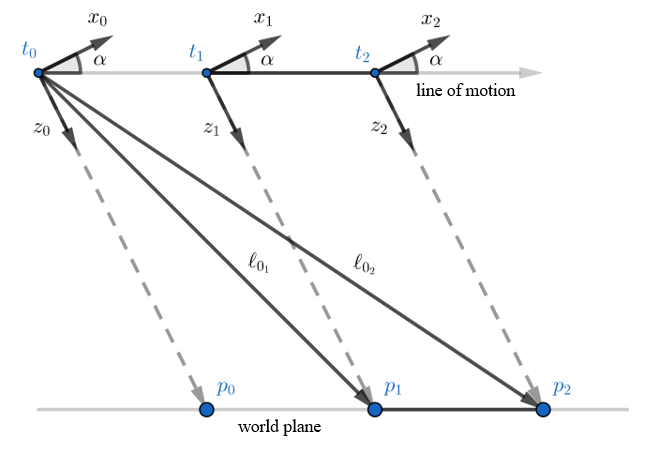}
    	\caption{Two points imaged at different time instants will have the same \(w\) only if the world plane and the camera motion are parallel.}
    	\label{fig:pushbroom_geometry}
    \end{figure}

    Next, consider two points observed at different times. Recall that the linear pushbroom model assumes that both the camera velocity and camera attitude are constant. It follows, therefore, that the ratio $V_z / V_x$ is constant. Next, observe that for $w$ to be constant from point to point means that it must be the same for two arbitrary points $\bp_1$ and $\bp_2$ residing on the conic. Therefore, as shown in Fig.~\ref{fig:pushbroom_geometry}, let $\bl_{0_1}$ and $\bl_{0_2}$ be the location of two points on the conic crater rim (and residing within the plane of the crater) relative to the camera's initial position. Substituting into Eq.~\eqref{eq:w_equation} and assuming $w$ is the same for both yields
    \begin{equation}
        \label{eq:w_equality_equation}
        -\ell_{0_{x_1}}\frac{V_z}{V_x} + \ell_{0_{z_1}} = w = -\ell_{0_{x_2}}\frac{V_z}{V_x} + \ell_{0_{z_2}}
    \end{equation}    
    This can be rearranged to find
    \begin{equation}
        \label{eq:velocity_slope_equality}
        \frac{V_z}{V_x}=\frac{\ell_{0_{z_2}}-\ell_{0_{z_1}}}{\ell_{0_{x_2}}-\ell_{0_{x_1}}} = \tan \alpha
    \end{equation}  
    where the angle $\alpha$ is shown in Fig.~\ref{fig:pushbroom_geometry}. This constraint states that the velocity vector must be parallel to the plane of the crater.
    
    Thus, the two constraints are that both the camera frame $y$-direction and the velocity vector are parallel to the plane of the crater. Since $V_x \neq 0$ if the pushbroom camera is to form a 2-D image, we know that the camera $y$-direction and velocity vector span a plane parallel to the crater plane. This same result may be arrived at by simplification of the implicit equation for a generic projection and is discussed in a later section.
    
\section{Generic Projection of a Conic into a Pushbroom Image}\label{sec:generic_projection_of_a_conic_into_a_pushbroom_image}
    
    We found in the prior section that an elliptical crater rim (or any other conic) will only project to another conic in a pushbroom image under very specific conditions. We now proceed to develop a more general description of the curve produced when a pushbroom camera images a conic feature. 
    
    Recalling the parametric expression of points on the conic given by Eq.~\eqref{eq:ParamXform}, we now substitute this into the pushbroom projection from Eq.~\eqref{eq:PushbroomMapping}:
    \begin{equation}
       \begin{bmatrix}
            u \\ vw \\w
        \end{bmatrix} = \frac{1}{\theta^2 + 1} \bM \bH \bar{\btheta}
    \end{equation}
    which expands to
    \begin{subequations}
    \begin{equation}
       u=\frac{A\theta^2+B\theta+C}{\theta^2+1}
    \end{equation}
    \begin{equation}
        vw=\frac{D\theta^2+E\theta+F}{\theta^2+1}
    \end{equation}
    \begin{equation}        
        w=\frac{G\theta^2+H\theta+I}{\theta^2+1}
    \end{equation}
    \end{subequations}
    The terms \(A, ..., I\) may be computed as 
    \begin{equation}\label{eq:MHmatrix}
        \bM\bH=\begin{bmatrix}
            A & B & C\\
            D & E & F\\
            G & H & I\\
        \end{bmatrix}
    \end{equation}
    where $\bM$ is from Eq.~\eqref{eq:DefM} and $\bH$ is from Eq.~\eqref{eq:HomographyForSingleParam}. The explicit expression of these coefficients can be found in Appendix~\ref{sec:appendix}. Finally, the parametric expression of the imaged curve is
    \begin{subequations}\label{eq:Explicit}
    \begin{equation}\label{eq:ucomponent}
        u=\frac{A\theta^2+B\theta+C}{\theta^2+1}
    \end{equation}
    \begin{equation}\label{eq:vcomponent}
        v=\frac{D\theta^2+E\theta+F}{G\theta^2+H\theta+I}
    \end{equation}
    \end{subequations}
    
    From these equations, we can derive the implicit equation of the apparent curve in the pushbroom image. From Eq.~\eqref{eq:ucomponent}, we can write
    \begin{equation*}
        (A-u)\theta^2+B\theta+(C-u)=0
    \end{equation*}
    and we can use it to express \(\theta^2\) as a rational function of \(u\) and \(\theta\)
    \begin{equation}
        \theta^2=\frac{(u-C)-B\theta}{(A-u)}
    \end{equation}
    Substituting this expression for \(\theta^2\) in Eq.~\eqref{eq:vcomponent} and manipulating, we obtain the expression for \(\theta\) as a function of \(u\) and \(v\)
    \begin{equation}\label{eq:thetaexplicit}
        \theta=\frac{(u-C)(vG-D)+(vI-F)(A-u)}{B(vG-D)+(E-vH)(A-u)}
    \end{equation}
    Substituting Eq.~\eqref{eq:thetaexplicit} into Eq.~\eqref{eq:ucomponent} and factoring out the term \((A-u)\), the final implicit expression of the curve can be written as a polynomial of degree four (a biquadric) of the form
    \begin{equation}\label{eq:ImplicitExpression}
        \alpha u^2v^2 + \beta u^2v +\gamma uv^2 + \delta uv + \epsilon u^2 +\zeta v^2 +\eta u + \iota v + \kappa=0
    \end{equation}
    where the coefficients \(\alpha, ..., \kappa\) are only dependent on \(A, ..., I\). The complete list of expressions for these coefficients is provided in Appendix~\ref{sec:appendix}. 
    
    Now, consider a point $\bxi^T \propto [uv, u, v, 1] \in \PP^1\times\PP^1$. In this case, the polynomial from Eq.~\eqref{eq:ImplicitExpression} can be written in compact form as
    \begin{equation}\label{eq:ImplicitCompact}
        \bar{\bxi}^T \bQ \bar{\bxi} = 0
    \end{equation}
    where $\bQ$ is the $4 \times 4$ matrix of arbitrary scale
    \begin{equation}
        \bQ \propto
        \begin{bmatrix}
            \alpha & \beta/2 & \gamma/2 & 0\\
            \beta/2 & \epsilon & \delta/2 & \eta/2 \\
            \gamma/2 & \delta/2 & \zeta & \iota/2\\
            0 & \eta/2 & \iota/2 & \kappa\\
        \end{bmatrix}
    \end{equation}
    From here, there are a few interesting analyses that may be performed. The first is to show (again) the conditions under which a world conic projects to a conic in a pushbroom image. The second is to discuss finding the projection as an intersection of two surfaces.

    \subsection{Reduction to a conic in the pushbroom image}\label{sec:reduction_to_a_pushbroom_conic}
    Consider the situation where we set the coefficients \(\alpha\), \(\beta\) and \(\gamma\) in Eq.~\eqref{eq:ImplicitExpression} to zero. In this case, $\bQ$ collapses to
    \begin{equation}
        \bQ_0 \propto
        \begin{bmatrix}
            0 & 0 & 0 & 0\\
            0 & \epsilon & \delta/2 & \eta/2 \\
            0 & \delta/2 & \zeta & \iota/2\\
            0 & \eta/2 & \iota/2 & \kappa\\
        \end{bmatrix}
        \propto \begin{bmatrix}
            0 & \textbf{0}_{1\times3}\\
            \textbf{0}_{3\times1} & \textbf{A}\\
        \end{bmatrix}
    \end{equation}
    where $\bA$ is a $3 \times 3$ matrix of arbitrary scale
    \begin{equation}
        \bA \propto
        \begin{bmatrix}
            \epsilon & \delta/2 & \eta/2 \\
            \delta/2 & \zeta & \iota/2\\
            \eta/2 & \iota/2 & \kappa\\
        \end{bmatrix}
    \end{equation}
    Recalling that $\bar{\bu}^T = [u,v,1] \in \PP^2$, it follows that
    \begin{equation}
        \label{eq:ConicPolynomialReduction}
        \bar{\bxi}^T \bQ_0 \bar{\bxi} = \bar{\bu}^T \bA \bar{\bu} = 0
    \end{equation}
    where $\bA$ is a constant matrix since it is made up of the constant coefficients from Eq.~\eqref{eq:ImplicitExpression}. Hence Eq.~\eqref{eq:ConicPolynomialReduction} describes a conic in $u$-$v$ coordinates (i.e., in the pushbroom image).

    Recalling from Table~\ref{tab:ImplicitCoefficients} that $\alpha=H^2+(G-I)^2$, we find the condition for \(\alpha=0\) is only satisfied when
    \begin{equation} \label{eq:HGIconstraint}
        H=0
        \quad \text{and} \quad
        G=I
    \end{equation}
    The reader may verify from Table~\ref{tab:ImplicitCoefficients} that this same condition also forces \(\beta=0\) and \(\gamma=0\). Imposing the two conditions of Eq.~\eqref{eq:HGIconstraint} is, therefore, necessary and sufficient to guarantee that the curve in the pushbroom image is a conic. 
    
    To better understand the meaning of Eq.~\eqref{eq:HGIconstraint}, consider the element-wise description of the rotation matrix \(\bT_C^M\) given by
    \begin{equation}\label{eq:Telements}
        \bT_C^M=
        \begin{bmatrix}
            \bt_x & \bt_y & \bt_z
        \end{bmatrix}
        =\begin{bmatrix}
        T_{11} & T_{12} & T_{13}\\
        T_{21} & T_{22} & T_{23}\\
        T_{31} & T_{32} & T_{33}\\
        \end{bmatrix}
    \end{equation}
    Using this notation, substitute the values for \(H\), \(G\) and \(I\) from Table~\ref{tab:ExplicitCoefficients} in the appendix into Eq.~\eqref{eq:HGIconstraint}. Then the two conditions can be rewritten as
    \begin{subequations}\label{eq:Tmatrixconditions}
    \begin{equation}\label{eq:TmatrixconditionY}
        H=0 \quad \rightarrow \quad \frac{T_{32}}{T_{12}}=\frac{V_z}{V_x}
    \end{equation}
    \begin{equation}\label{eq:TmatrixconditionX}
        G=I \quad \rightarrow \quad \frac{T_{31}}{T_{11}}=\frac{V_z}{V_x}
    \end{equation}
    \end{subequations}
    
    We can show that this pair of constraints for a world conic to project to a pushbroom conic is the same as found in earlier sections---namely that the velocity vector and camera frame $Y$-axis span a plane parallel to the plane of the crater.
    This requires a few short steps. First, the normal to the plane spanned by the velocity vector and camera frame $Y$-axis is in the direction given by
    \begin{equation}    
        \bv \times \hat{\by} = 
        \begin{bmatrix}
            V_x \\ V_y \\ V_z
        \end{bmatrix} \times
        \begin{bmatrix}
            0 \\ 1 \\ 0
        \end{bmatrix} = 
        \begin{bmatrix}
            -V_z \\ 0 \\ V_x
        \end{bmatrix}
    \end{equation}
    and normal to the crater plane (the Moon $X$-$Y$ plane) is in the direction given by
    \begin{equation}    
        \bt_x \times \bt_y = 
        \begin{bmatrix}
            T_{21}T_{32} - T_{22}T_{31} \\ 
            T_{12}T_{31} - T_{11}T_{32} \\ 
            T_{11}T_{22} - T_{12}T_{21}
        \end{bmatrix}
    \end{equation}
    If the planes are parallel then their normals are also parallel,
    \begin{equation} 
        \begin{bmatrix}
            -V_z \\ 0 \\ V_x
        \end{bmatrix} = \bv \times \hat{\by}
        \propto
        \bt_x \times \bt_y = 
        \begin{bmatrix}
            T_{21}T_{32} - T_{22}T_{31} \\ 
            T_{12}T_{31} - T_{11}T_{32} \\ 
            T_{11}T_{22} - T_{12}T_{21}
        \end{bmatrix} 
    \end{equation}
    Since the second row is zero (i.e., the normal vector must lie in the camera frame $x$-$z$ plane) then
    \begin{equation}    
            T_{12}T_{31} - T_{11}T_{32} = 0 
    \end{equation}
    which certainly agrees with Eq.~\eqref{eq:Tmatrixconditions}:
    \begin{equation}
        \frac{T_{32}}{T_{12}}=\frac{V_z}{V_x} = 
        \frac{T_{31}}{T_{11}} \rightarrow T_{12}T_{31} = T_{11} T_{32}
    \end{equation}
    However, more can be done to separately arrive at the two equations in Eq.~\eqref{eq:Tmatrixconditions}. If $\bv, \hat{\by}, \bt_x, \bt_y$ are all coplanar then their cross products are all parallel. For example,
    \begin{equation} 
        \bv \times \hat{\by}
        \propto
        \bt_x \times \bt_y 
        \propto
        \bt_x \times \bv 
        \propto
        \bt_y \times \bv 
    \end{equation}
    It follows, therefore, that
    \begin{equation}    
        \bt_x \times \bv = 
        \begin{bmatrix}
            T_{21}V_z - T_{31}V_y  \\ 
            T_{31}V_x - T_{11}V_z \\ 
            T_{11}V_y - T_{21}V_x
        \end{bmatrix} \propto 
        \begin{bmatrix}
            -V_z \\ 0 \\ V_x
        \end{bmatrix}
    \end{equation}
    And so, from the second row, we directly reproduce Eq.~\eqref{eq:TmatrixconditionX}
    \begin{equation}
        T_{11}V_z - T_{31}V_x = 0 \rightarrow \frac{T_{31}}{T_{11}} = \frac{V_z}{V_x}
    \end{equation}
    Likewise, we see that
    \begin{equation}    
        \bt_y \times \bv =
        \begin{bmatrix}
            T_{22}V_z -T_{32}V_y \\ 
            T_{32}V_x - T_{12}V_z \\ 
            T_{12}V_y -T_{22}V_x
        \end{bmatrix} \propto
        \begin{bmatrix}
            -V_z \\ 0 \\ V_x
        \end{bmatrix}
    \end{equation}
    where the second row now directly reproduces Eq.~\eqref{eq:TmatrixconditionY}
    \begin{equation}
         T_{32}V_x - T_{12}V_z = 0 \rightarrow \frac{V_z}{V_x} = \frac{T_{32}}{T_{12}}
    \end{equation}
    Thus, we see that Eq.~\eqref{eq:Tmatrixconditions} describes the situation where the velocity vector $\bv$ and camera frame $y$-direction span a plane parallel to the crater plane (which is spanned by $\bt_x$ and $\bt_y$).

    \begin{figure}[h]
    	\centering
    	\includegraphics[width=0.5\textwidth]{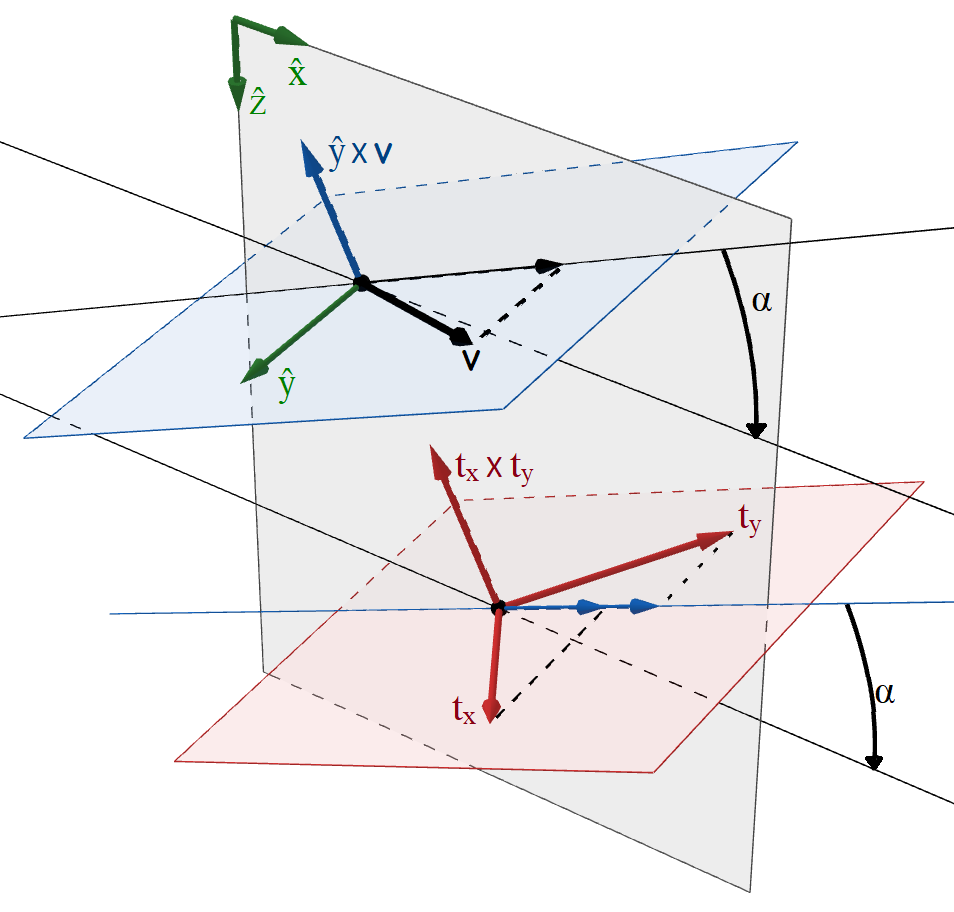}
    	\caption{When the plane of the ellipse (in red) and the plane spanned by the velocity vector and the $Y$-axis of the camera frame (in blue) have the same normal vector, the projection of the conic in the world frame is a conic in the pushbroom image.}
    	\label{fig:LastParSpecialCase}
    \end{figure}
    
    \subsection{Generic conic projection as the intersection of curves}\label{sec:generic_pushbroom_conic_projection_as_the_intersection_of_curves}
    We now return to studying the general conic projection described by the curve of Eq.~\eqref{eq:ImplicitCompact}. First, note that since the matrix \(\textbf{Q}\) is \(4\times4\) and symmetric, we can always associate it to a quadric in \(\PP^3\) described by the equation
    \begin{equation}
    \bar{\bzeta}^T
       \textbf{Q} \bar{\bzeta} = 0
    \end{equation}
    where $\bar{\bzeta} = [z,u,v,1] \in \PP^3$.
    The curve traced by the pushbroom camera can be formed by the intersection of this quadric with the surface \(z=uv\) (i.e., the constraint on the first component of $\bar{\bzeta}$ to get $\bar{\bzeta}=\bar{\bxi}$). Orthographic projection of this 3-D curve down to the \(u\)-\(v\) plane provides the 2-D curve we seek.

\begin{figure}[h]
    	\centering
    	\includegraphics[width=0.7\textwidth]{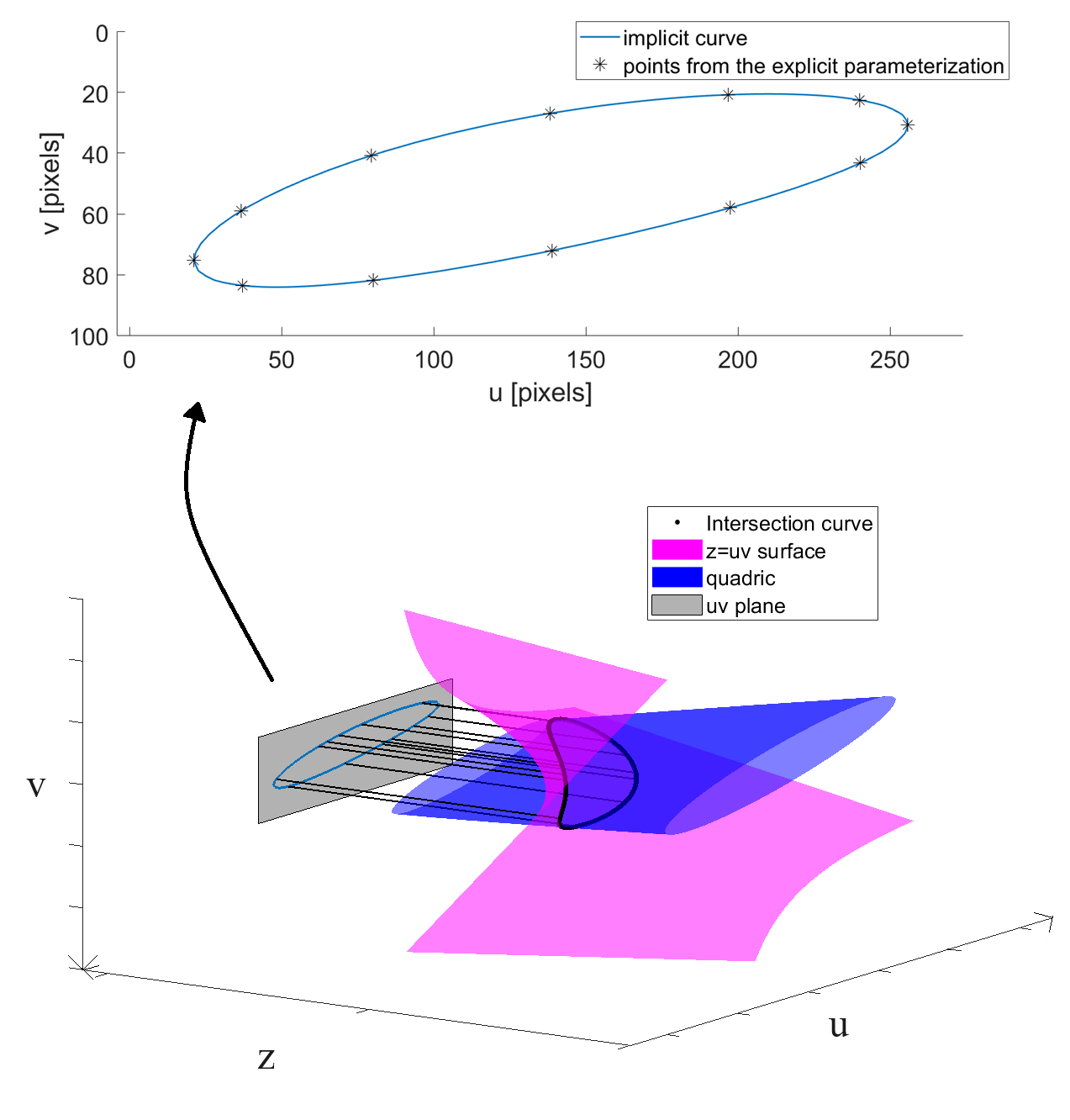}
    	\caption{Interpretation of the implicit equation of the projected curve. The projection of the intersection between the quadric and the surface \(z=uv\) perfectly matches points sampled from the explicit parameterization of the curve.}
    	\label{fig:Intersections}
    \end{figure}

    To show this relationship, consider the 3-D scenario illustrated in Fig.~\ref{fig:PushbroomDistortions}. The conic projection may be be found by intersecting the quadric surface~$\bQ$ and the surface \(z=uv\), as shown by the black curve in Fig.~\ref{fig:Intersections}. This is then orthographically projected onto the \(u\)-\(v\) plane. The agreement between the projected curve and the explicit representation of Eqs.~\eqref{eq:Explicit} is shown using \(\theta\) given by Eq.~\eqref{eq:thetaPhiRelation} for values of \(\phi\in(0,\,2\pi)\) sampled at intervals of \(30\) deg. This provides a way to interpret the implicit equation of the fourth-degree polynomial curve given by the projection of a conic provided by a linear pushbroom camera, and shows that it is equivalent to the explicit parameterization of the curve in Eqs.~\eqref{eq:Explicit}.
    
\section{Estimating the Curve on the Pushbroom Image}\label{sec:estimating_the_curve}
Suppose we measure points with coordinates $(u_i,v_i)$ along the rim of the crater in a pushbroom image. If we know the dimensions of the crater and the attitude of the camera with respect to the lunar frame \(\bT_C^M\), we may compute the \(\alpha,...,\kappa\) coefficients of the polynomial curve of Eq.~\eqref{eq:ImplicitExpression}. To do that, it is convenient to work in the image plane coordinates, which may be obtained from the pixel coordinates by inverting Eq.~\eqref{eq:uv1TOxy1}. For each pair \((u_i,v_i)\), we can determine
\begin{equation}\label{eq:pixelToimage}
\begin{bmatrix}
x_i\\
y_i\\
1
\end{bmatrix} = \begin{bmatrix}
    \tau & 0 & 0\\
    0 & 1/d_y & -v_p/d_y\\
    0 & 0 & 1
\end{bmatrix}\begin{bmatrix}
    u_i\\
    v_i\\
    1
\end{bmatrix}
\end{equation}
The coefficients \(\hat{\alpha},...,\hat{\kappa}\) corresponding to the curve in the image plane coordinates \(\left(x,\,y\right)\) can be estimated by solving the following least-squares problem
\begin{equation}\label{eq:least_square_problem}
    \begin{bmatrix}
        x_1^2 y_1^2 & x_1^2 y_1 & x_1 y_1^2 & x_1 y_1 & x_1^2 & y_1^2 & x_1 & y_1 & 1 \\
        x_2^2 y_2^2 & x_2^2 y_2 & x_2 y_2^2 & x_2 y_2 & x_2^2 & y_2^2 & x_2 & y_2 & 1 \\
        \vdots & & & \vdots & &\vdots & & & \vdots\\
        x_N^2 y_N^2 & x_N^2 y_N & x_N y_N^2 & x_N y_N & x_N^2 & y_N^2 & x_N & y_N & 1
    \end{bmatrix}
     \begin{bmatrix}
        \hat{\alpha} \\ \hat{\beta} \\ \hat{\gamma} \\ \hat{\delta} \\ \hat{\epsilon} \\ \hat{\zeta} \\ \hat{\eta} \\ \hat{\iota} \\ \hat{\kappa}
    \end{bmatrix} = \textbf{0}_{N \times 1}
\end{equation}
A minimum of eight crater rim points are required to ensure the least-squares problem is well-posed. Note that the coefficients \(\hat{\alpha},...,\hat{\kappa}\) have ambiguous scaling, so we need to work with their ratios. After the least-squares problem has been solved, we chose to normalize all the coefficients with respect to \(\hat{\epsilon}\) to simplify the following calculations. We will show now how the correct $\hat{\epsilon}$ can be computed to resolve the coefficients' scale ambiguity. 

Let \(\bar{\alpha},...,\bar{\kappa}\) be the normalized coefficients (note that \(\bar{\epsilon}=1\)), and define the following intermediate variables: 
\begin{equation}\label{eq:v_prime_def}
    \begin{bmatrix}
    v_1'\\
    v_2'\\
    v_3'\\
    \end{bmatrix}=\frac{1}{V_x}\begin{bmatrix}
    1\\
    V_y\\
    V_z\\
    \end{bmatrix}
\end{equation}
Also, let
\begin{equation}\label{eq:K}
    \begin{bmatrix}
    K_1 & K_2 & K_3\\
    K_2 & K_4 & K_5\\
    K_3 & K_5 & K_6\\
    \end{bmatrix}=a^2\bt_x\bt_x^T+b^2\bt_y\bt_y^T
\end{equation}

\begin{equation}\label{eq:m}
    \bm=ab\bt_z
\end{equation}
and note that the elements \(K_i\) and \(\bm\) only depend on known quantities. 

Using the expressions for \(\bar{\alpha}\) and \(\bar{\beta}\) derived from \(\hat{\alpha}\), \(\hat{\beta}\) and \(\hat{\epsilon}\) in Table \ref{tab:ImplicitCoefficients}, we can analytically solve for \(v_3'\) and \(v_2'\):
\begin{equation}
    v_3'=\frac{1}{K_1}\left[K_3\pm{\left(-m_2+2\frac{\bar{\alpha}}{\bar{\beta}}m_3\right)}\sqrt{\frac{\bar{\beta}^2}{4\bar{\alpha}-\bar{\beta}^2}}\right]
\end{equation}
\begin{equation}
    v_2'=\frac{K_1 v_3'^2 -2(K_3+K_2\frac{\bar{\alpha}}{\bar{\beta}})v_3'+ K_6 + 2 K_5 \frac{\bar{\alpha}}{\bar{\beta}} }{2\frac{\bar{\alpha}}{\bar{\beta}}(K_3-K_1v_3')}
\end{equation}
and for the scale factor \(\hat{\epsilon}\):
\begin{equation}
    \hat{\epsilon}=4 K_1 v_2'^2 - 8 K_2 v_2' + 4 K_4
\end{equation}
This solves the scale ambiguity:
\begin{equation}
\begin{bmatrix}
\hat{\alpha}\\
\vdots\\
\hat{\kappa}\\
\end{bmatrix}=\hat{\epsilon}\begin{bmatrix}
\bar{\alpha}\\
\vdots\\
\bar{\kappa}\\
\end{bmatrix}
\end{equation}

Given at least eight sampled points on the crater's rim, this process estimates the curve around the crater as seen in the pushbroom image. In theory, it also allows for an analytical solution of the position and velocity of the spacecraft in terms of the coefficients \(\hat{\alpha},\,...,\,\hat{\kappa}\). However, poor observability of some of the coefficients make the corresponding relative error significant when the measurements are affected by noise (as is the case in real applications). Even if the difference between the true and the estimated curve cannot be visually appreciated, these errors can have a major impact on the accuracy of the spacecraft state estimate. For this reason, in the next section we will introduce a solution to the problem of state estimation that departs from the procedure just described. 

\section{Motion from Pushbroom Images}\label{sec:motion_from_pushbroom_images}
Assume that our spacecraft is flying over a known crater (e.g., after crater identification has already been performed \cite{Christian:2021}), with known attitude (e.g., provided by star-trackers), but with unknown position and velocity. The pushbroom camera projects the elliptical crater on the pushbroom image as a curve which is a polynomial of degree four. In principle, if we determine the coefficients \(\hat{\alpha},...,\hat{\kappa}\) solving the least-square problem in the previous section, we can use them to obtain an analytical expression for the position and velocity of the camera capturing the image. As already mentioned at the end of Sec.~\ref{sec:estimating_the_curve}, however, the smallest amount of noise makes the estimate of some of the coefficients unreliable for a quantitative analysis. For this reason, we approached the problem differently. In the rest of this section, we will show that a set of \(N\geq 8\) pixel coordinates pairs \((u_i,v_i)\) may be used to determine the initial position and velocity of the camera by introducing a least-squares problem formulated as a polynomial system, that can be solved via homotopy continuation \cite{Sommese}. We then apply this method to recover the position and velocity of a camera from a real pushbroom image.

\subsection{Polynomial formulation of the least-squares problem}
Consider the pushbroom image of a crater and a point \(\left(u_i,\,v_i\right)\) on its rim. For convenience, the pixel coordinates are transformed into image plane coordinates \({\bx}_i^T\propto \left[ x_i\,y_i\,1 \right]\) using Eq.~\eqref{eq:pixelToimage} so that
\begin{equation}\label{eq:implicit_compact_with_hat}
   {\bx}_i^T \boldsymbol{\hat{Q}}{\bx}_i=0
\end{equation}
Note that Eq.~\eqref{eq:implicit_compact_with_hat} is a polynomial in the unknowns \(\bX^T=\left[\brr_{0,C}^T\quad\bv'^T\right]\), where we recall that \(\brr_{0,C}=\bT_C^M \brr_0\) is the initial position of the camera relative to the crater's center expressed in the camera frame, and the components of \(\bv'\) are defined in Eq.~\eqref{eq:v_prime_def}. The explicit expression of the coefficients of \(\boldsymbol{\hat{Q}}=\boldsymbol{\hat{Q}}(\bX)\) is provided in Appendix~\ref{sec:appendix}.

Given the coordinates of \(N\) points on the pushbroom image, we can formulate the cost function
\begin{equation}\label{eq:J}
    J = \sum_{i=1}^N \left(\bx_i^T \hat{\bQ}\bx_i\right)^2
\end{equation}
where \(J=0\) when \(\boldsymbol{\hat{Q}}\) is calculated using the true state \(\bX^*\). Therefore, the actual state is a minimizer of the cost function \(J\). 

The differentiation of \(J\) with respect to the elements of \(\bX\) produces six equations, that are polynomials in \(\bX\):
\begin{equation}\label{eq:poly_cost}
    \nabla J_{\bX}=\textbf{0}_{6\times 1}
\end{equation}
The solutions to Eq.~\eqref{eq:poly_cost} provide all the critical points of the cost function~\(J\). Once these critical points have been found, only those that produce values of the cost function \(J(\bX)\) equal to zero (with ideal measurements) or close to zero (with real measurements) should be considered. 

We chose to numerically solve the polynomial system of Eq.~\eqref{eq:poly_cost} via homotopy continuation. This choice has the advantage of providing the solutions of the polynomial system without requiring an initial guess of the state \(\bX\). Details on the specific homotopy used are given in the next section. 

\subsection{Homotopy continuation}
Consider a square \textit{target system} \(\mathcal{F}\) that we wish to solve, and another square \textit{start system} \(\mathcal{G}\) whose solutions are known. From \(\mathcal{F}\) and \(\mathcal{G}\), we can build the auxiliary system \(\mathcal{H}\) :
\begin{equation}\label{eq:genericHomotopy}
    \mathcal{H}(t_{H}) = (1-t_{H}) \mathcal{G} + t_{H} \mathcal{H}
\end{equation}
which depends on the continuation parameter \(t_{H}\). When \(t_{H}=0\), the solutions of \(\mathcal{H}\) coincide with the solutions of the start system, and are known. When \(t_{H}=1\), the solutions of the system \(\mathcal{H}\) coincide with the solutions of the target system that we wish to estimate. 
A homotopy continuation method is a predictor-corrector method that tracks the solutions of \(\mathcal{H}\) as the continuation parameter \(t_H\) is varied from \(0\) to \(1\).

From each solution of \(\mathcal{G}\) originates a path that the numerical method tracks and that potentially leads to a solution of \(\mathcal{F}\). If the number of solutions of \(\mathcal{G}\) is smaller than the number of solutions of \(\mathcal{F}\), then some solutions of the target system will be lost. On the contrary, if \(\mathcal{G}\) has more solutions than \(\mathcal{F}\), some paths will be divergent and computationally expensive to track. In the ideal case, \(\mathcal{G}\) and \(\mathcal{F}\) have the same number of solutions. 

\subsubsection{Parameter homotopy}\label{sec:parameterHomotopy}
\begin{figure}[ht!]
    \centering
    \includegraphics[width=0.6\textwidth]{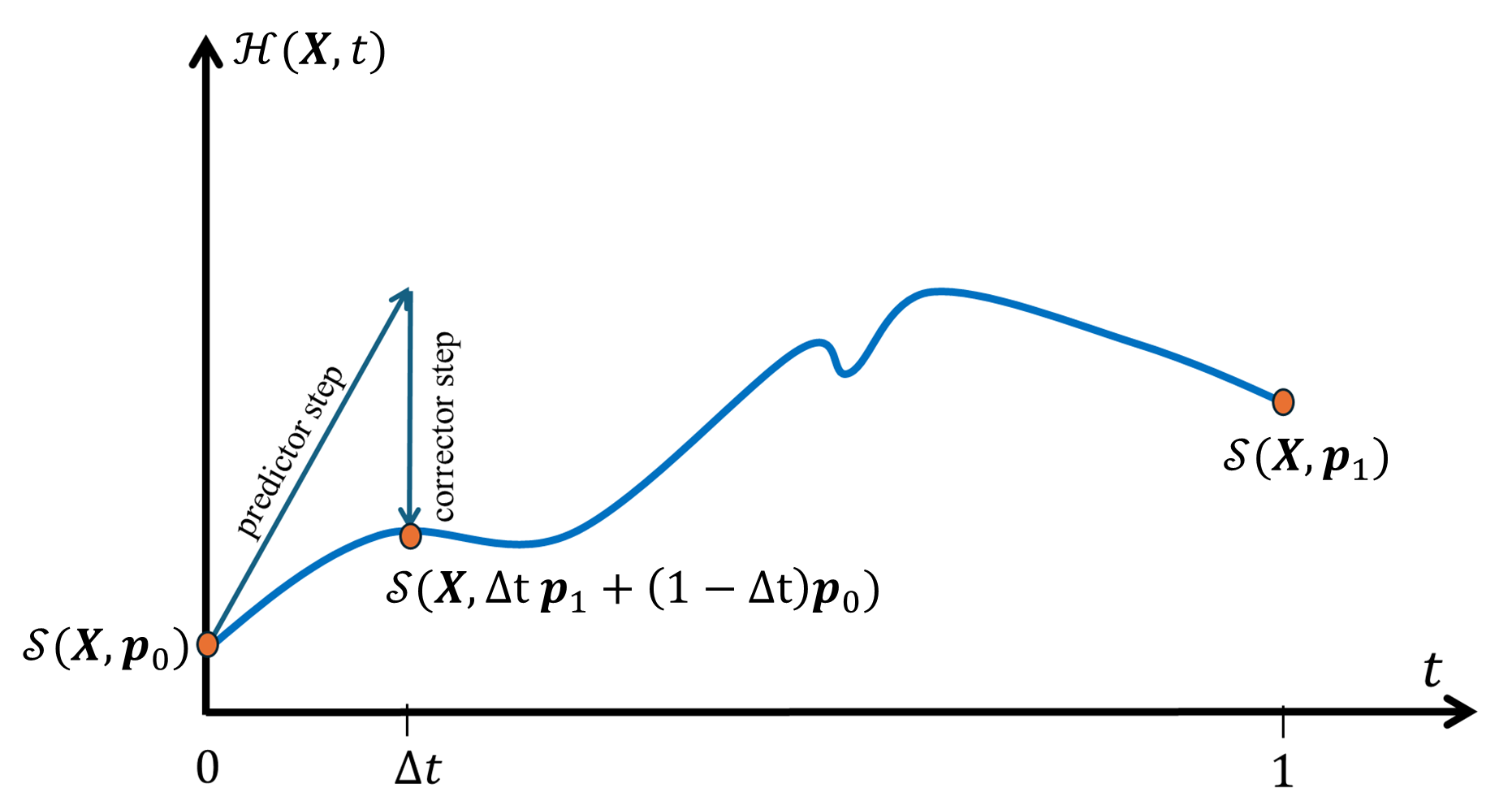}
    \caption{Starting from the solution at \(t=0\), a predictor method (e.g. Runge Kutta) predicts the solution at the following time step \(t=\Delta t\). A corrector method (e.g. Newton's method) is then used to refine this initial guess, providing the refined solution of the system \(\mathcal{H}(\bX,\Delta t)\).}
    \label{fig:homotopy}
\end{figure}

Consider a generic polynomial system \(\mathcal{S}\) of \(N\) equations in \(N\) unknowns. Let \(\bX\) be the vector of unknowns and let \(\bp\) be a vector of parameters. When the parameters are set, they uniquely identify the coefficients of the polynomial equations inside \(\mathcal{S}\). Now assume that our target system is identified by a specific set of parameters \(\bp_1\). In other words, let \(\mathcal{F}\) be compactly defined as: 
\begin{equation}
    \mathcal{F}(\bX)=\mathcal{S}(\bX,\bp_1)
\end{equation}
Furthermore, assume that our start system is characterized by another given set of parameters \(\bp_0\):
\begin{equation}
    \mathcal{G}(\bX)=\mathcal{S}(\bX,\bp_0)
\end{equation}
Here, the auxiliary system \(\mathcal{H}\) of Eq. \eqref{eq:genericHomotopy} can be re-written as
\begin{equation}\label{eq:parameterHomotopy}
    \mathcal{H}(\bX,t)=\mathcal{S}\left(\bX,(1-t_H)\bp_0+t_H\bp_1\right)
\end{equation}
In this case, changing \(t_H\) only affects the parameters (and consequently the coefficients) of the polynomial equations. This type of homotopy is called \textit{parameter homotopy}, and a 1-D schematic representation is shown Fig. \ref{fig:homotopy}.

Parameter homotopy has the advantage of providing a start system and a target system which have the same exact number of solutions, minimizing the number of paths that need to be tracked. For additional details on parameter homotopy, refer to Ref. \cite{Morgan:1989}. 

In our specific application, the target system \(\mathcal{F}\) is the cost function of Eq.~\eqref{eq:poly_cost}. The choice of the parameters may be done in multiple ways, and will be discussed below. We built the start system by randomly selecting the parameter \(\bp_0\) in the complex space, and determined its solution using a technique based on monodromy \cite{Duff:2019,Duff:2021,Amendola:2016}. If the parameters \(\bp_0\) are sufficiently generic, then the solutions to the start system can be stored and re-used to solve any problem of the same class --- i.e. any other pushbroom camera's state estimation problem. The path tracking was done using an implementation of parameter homotopy provided by the software package \texttt{NAG4M2} in the computer algebra system \texttt{Macaulay2} \cite{macaulay2}. 

\subsubsection{Implementation details}
Implementing the solver in \texttt{Macaulay2} requires setting up the polynomial system (the \(\mathcal{S}\) of Sec.~\ref{sec:parameterHomotopy}) and declaring the parameters and variables of the problem. The choices made for this study and their rationals are discussed here. 

There are different possible choices for the parameters of the system. If we fix the number \(N\) of points on the pushbroom image, then the parameters can be set as the elements of the attitude matrix \(\bT_C^M\), the crater's dimensions \(a\) and \(b\), and the coordinates of the points \((u_i,\,v_i)\), for a total of \(11+2N\) parameters. In this way, we can track exactly the optimal number of paths for a given \(N\). However, this approach has a major downside: the number of solutions can change if the number of parameters changes. This will require building a new start system every time the selected number of points changes, which is undesirable. 

We can instead choose the parameters as the coefficients of the polynomial expression of \(J\) in Eq.~\eqref{eq:J}. This produces a standardized expression of the system of Eq.~\eqref{eq:poly_cost} that can be used with any \(N\). The total number of paths tracked with this approach is 243, up to a sign symmetry. In fact, if \(\bX\) is a solution to the problem, the same holds for \(-\bX\). This number of paths will be slightly sub-optimal for a fixed number of points \(N\), so the actual number of independent solutions will be smaller. However, for comparison, we can determine the number of paths that the total degree homotopy, a more common choice, would require to track. This number can be found by applying Bézout's theorem \cite{Harris:1992}, which in this case gives \(7^4\cdot6^2>>243\) paths.

\subsection{Identification of the true solution}
At the end of the tracking, we can determine the unique true solution by exploiting the knowledge of the direction of motion and a few simple physical considerations, described in this section. Once the 243 solutions are known, we can eliminate those that are not physically possible. First, we know that the camera flies \textit{above} the ground, and therefore the \(z\)-component of the position vector expressed in the camera frame must be negative. Because of the sign symmetry, we can replace any solution that has a positive \(z\) component with its opposite. Since the direction of motion fixes the sign that \(V_x\) has in the camera frame, we can further discard all solutions that produce a sign of \(V_x\) that disagrees with the direction of motion. The remaining candidates are then ordered based on the value of the cost function \(J\), given by Eq.~\eqref{eq:J}, and we take the candidate that produces the minimum cost. In all of our experiments, this solution identification approach always yielded the correct solution.

\section{Numerical Results}\label{sec:numerical_results}
The objective of this section is to test the accuracy of the projection model described in this work. For these experiments, we used pushbroom images captured by the Lunar Reconnaissance Orbiter's NAC \cite{Robinson:2010}.  We will first qualitatively test the accuracy of the rim modeling given by Eq.~\eqref{eq:ImplicitCompact}. Afterwards, we will use points on the crater rims in the images to recover the state of the spacecraft with the methodology described in Sec.~\ref{sec:motion_from_pushbroom_images}. 

\subsection{Projection on the pushbroom image}
    For our experimental validation, we selected LROC NAC images from \cite{Robinson:2010} which are catalogued in NASA's Planetary Data System (PDS)\cite{Robinson:2010b} and registered with NAIF SPICE kernels~\cite{Bowman-Cisneros:2010}. Images can also be selected from Ref. \cite{QuickMap}, a tool that allows the user to directly select images containing the desired craters. 
    From the above mentioned kernels, we retrieved the camera intrinsic parameter (i.e. focal length, pixel pitch, principal point and exposure time), as well as the camera position, velocity and attitude at any time during the image capture. We retrieved the crater dimensions and orientation from the Robbins crater database, and the crater center from Ref. \cite{QuickMap}. This choice was motivated by the desire of limiting the differences between the lunar frame utilized to define the crater center and the NAIF kernels. 
    
    In general, the spacecraft motion with respect to the surface is non-linear, however, under a linear pushbroom model, we assume the camera velocity and attitude are constant. For each crater, we linearize the trajectory when the instantaneous view plane crosses the crater center $\ell_x=0$ which is easily determined in the camera frame. The time from the start of the image to this point is $\Delta t^*$, and the camera position and velocity are $\brr^*$ and $\bv^*$. Now, the velocity is constant ($\bv_0=\bv^*$) and the initial position is $\brr_0=\brr^*-\Delta t^* \bv^*$ via Eq.~\eqref{eq:position_constant_velocity_kinematics}.

\begin{figure}
\subfloat[Curtis (right NAC), LROC Image \texttt{M1107903421RE} ]{\includegraphics[width = \textwidth]{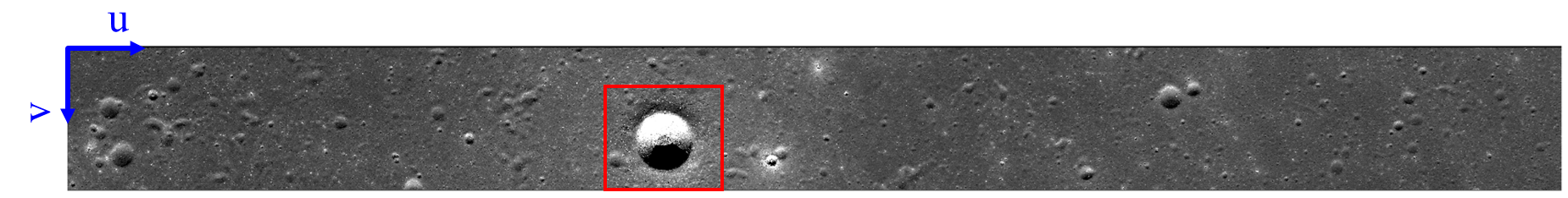}} \\
\subfloat[Palisa A (right NAC), LROC Image \texttt{M102200777RE} ]{\includegraphics[width = \textwidth]{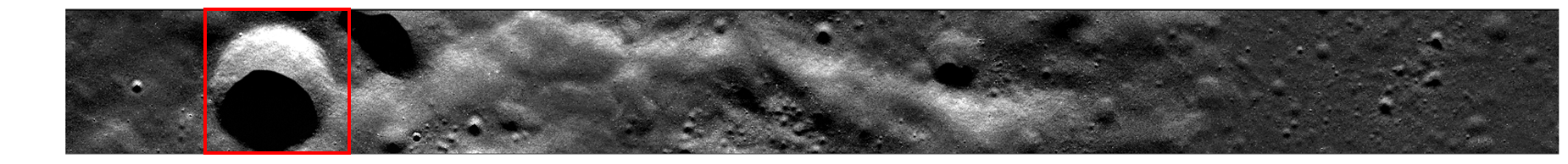}}\\
\subfloat[Curtis (left NAC), LROC Image \texttt{M1166793617LE}]{\includegraphics[width = \textwidth]{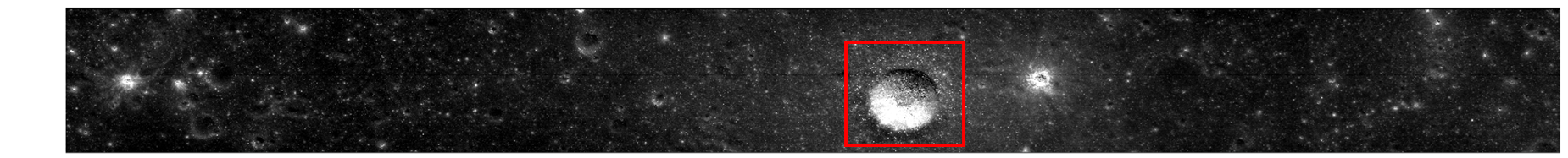}}\\
\subfloat[Euler L (right NAC), LROC Image \texttt{M1213270577RE} ]{\includegraphics[width = \textwidth]{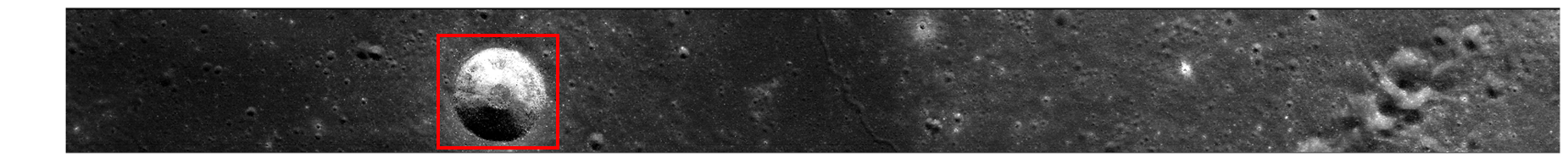}} \\
\subfloat[Hortensius (right NAC), LROC Image \texttt{M1112040045RE} ]{\includegraphics[width = \textwidth]{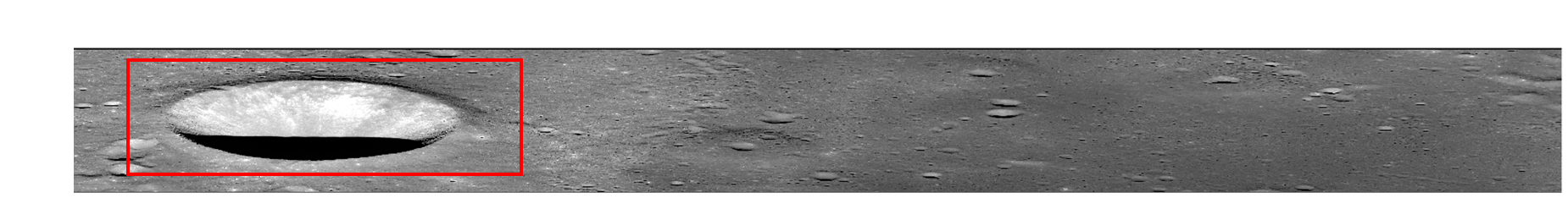}} \\
\caption{Five pushbroom images captured by the LRO NAC cameras. The size in pixels of each image is \(52225\, \times 5065\). See Fig.~\ref{fig:craters_and_curve} for a zoom of the highlighted regions.}
\label{fig:all_full}
\end{figure}

\begin{figure}
\centering
\subfloat[Curtis (right NAC)]{\includegraphics[width = 0.4\textwidth]{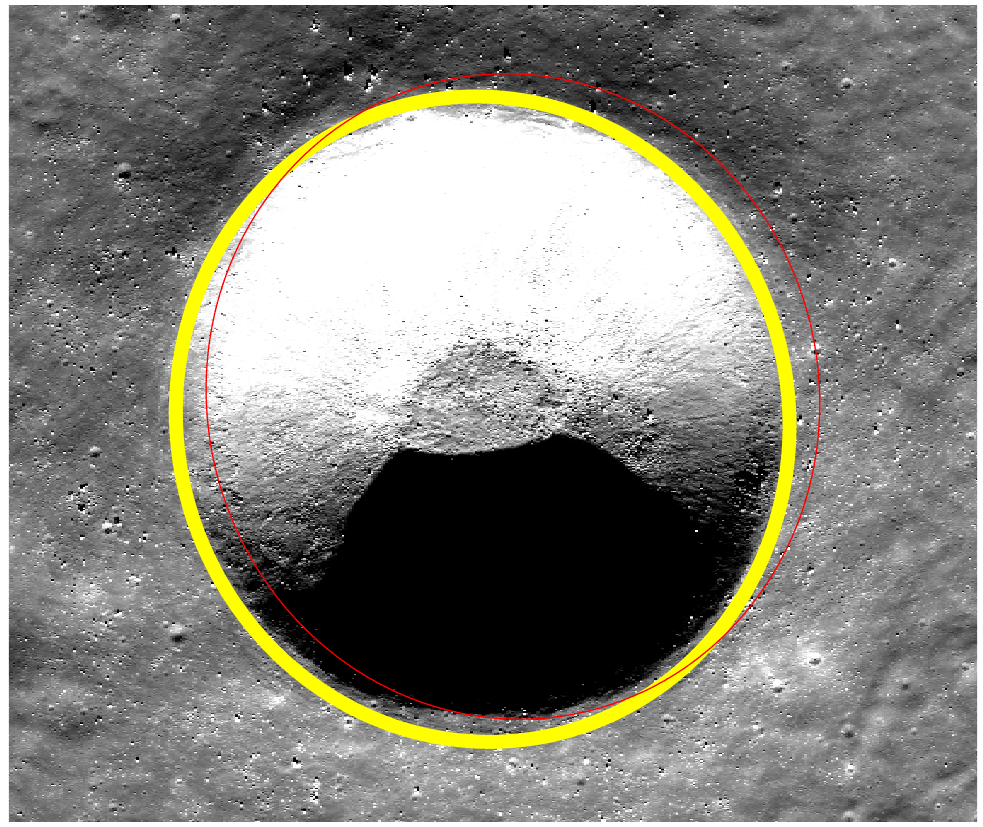}} 
\hspace{0.05cm}
\subfloat[Palisa A (right NAC)]{\includegraphics[width = 0.4\textwidth]{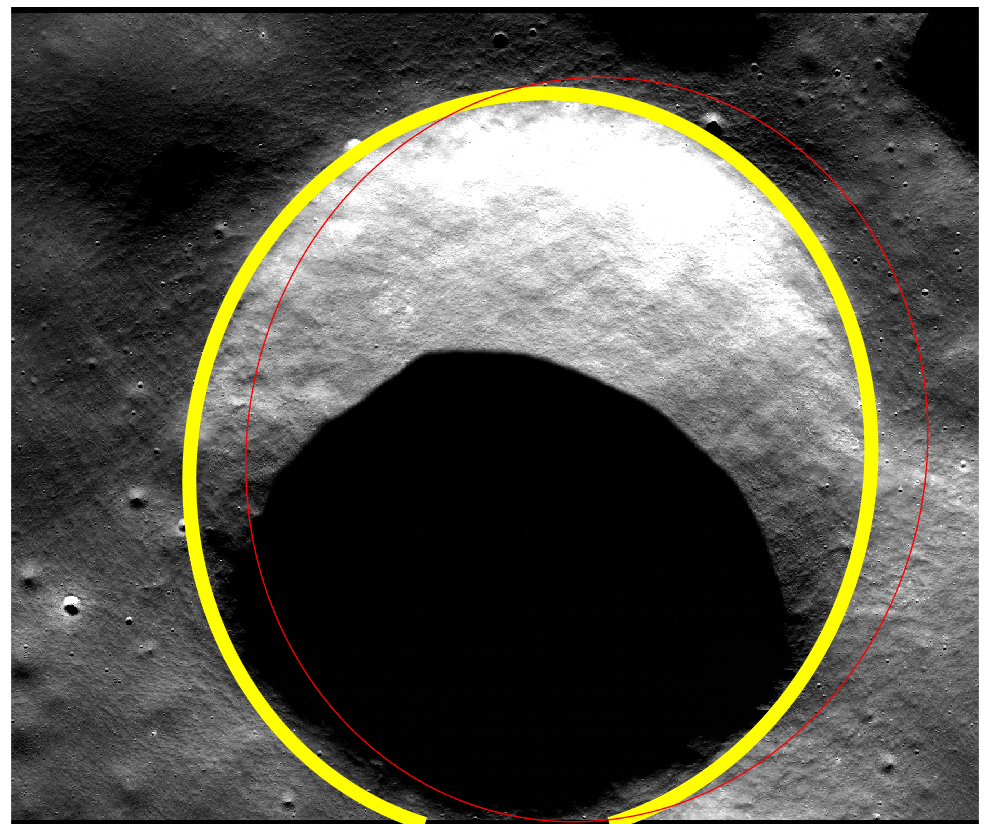}}\\
\subfloat[Curtis (left NAC)]{\includegraphics[width = 0.4\textwidth]{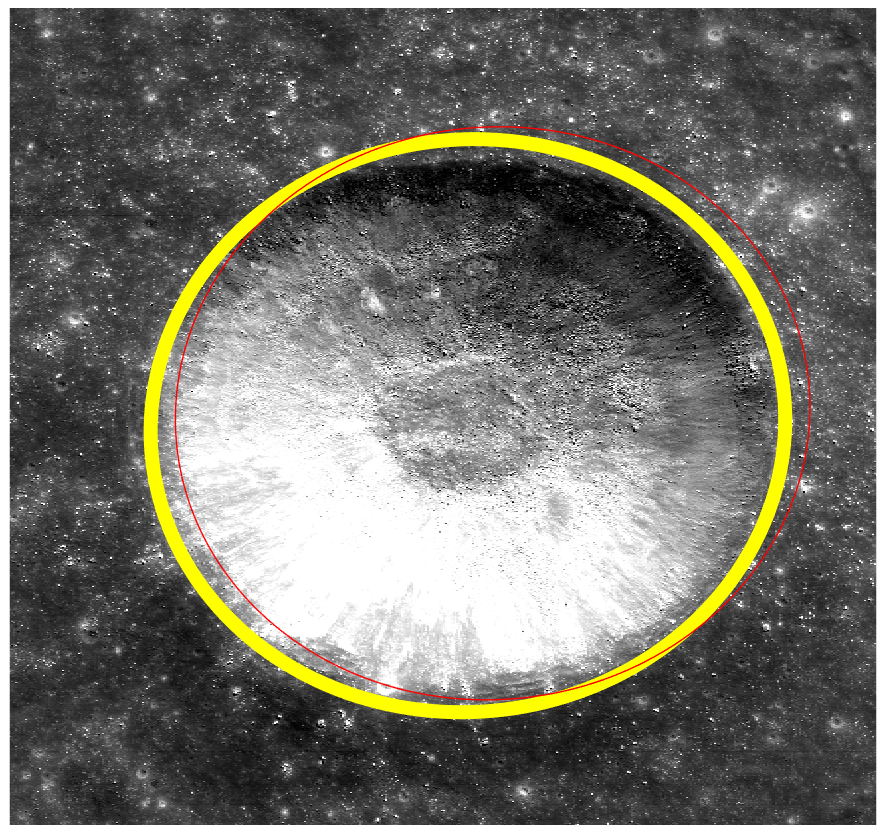}}
\hspace{0.05cm}
\subfloat[Euler (right NAC)]{\includegraphics[width = 0.4\textwidth]{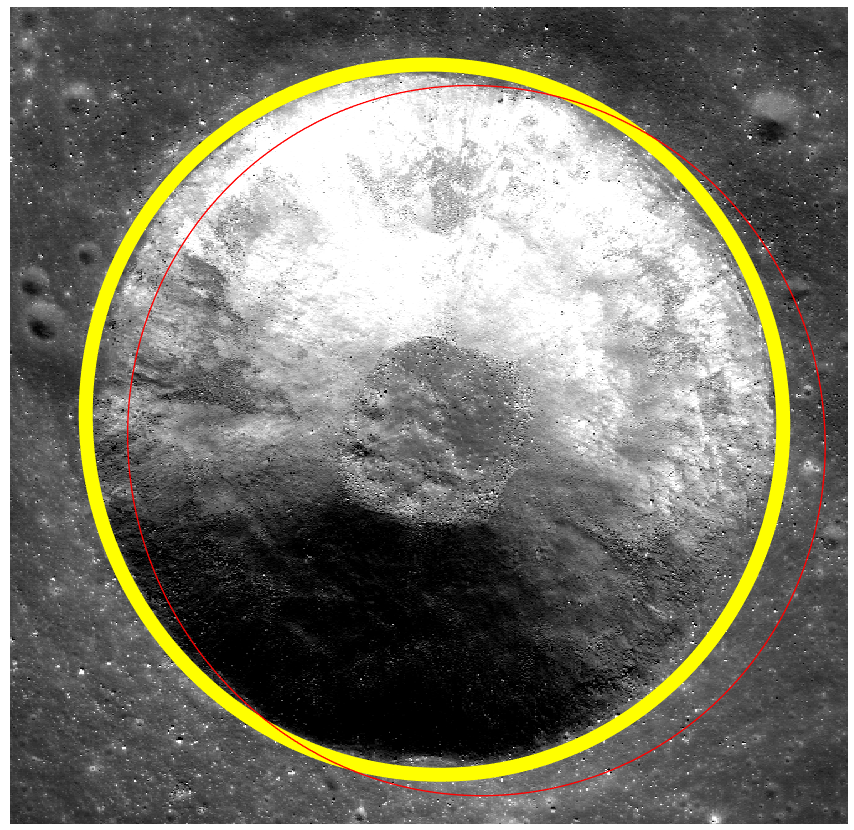}} \\
\subfloat[Hortensius (right NAC)]{\includegraphics[width = \textwidth]{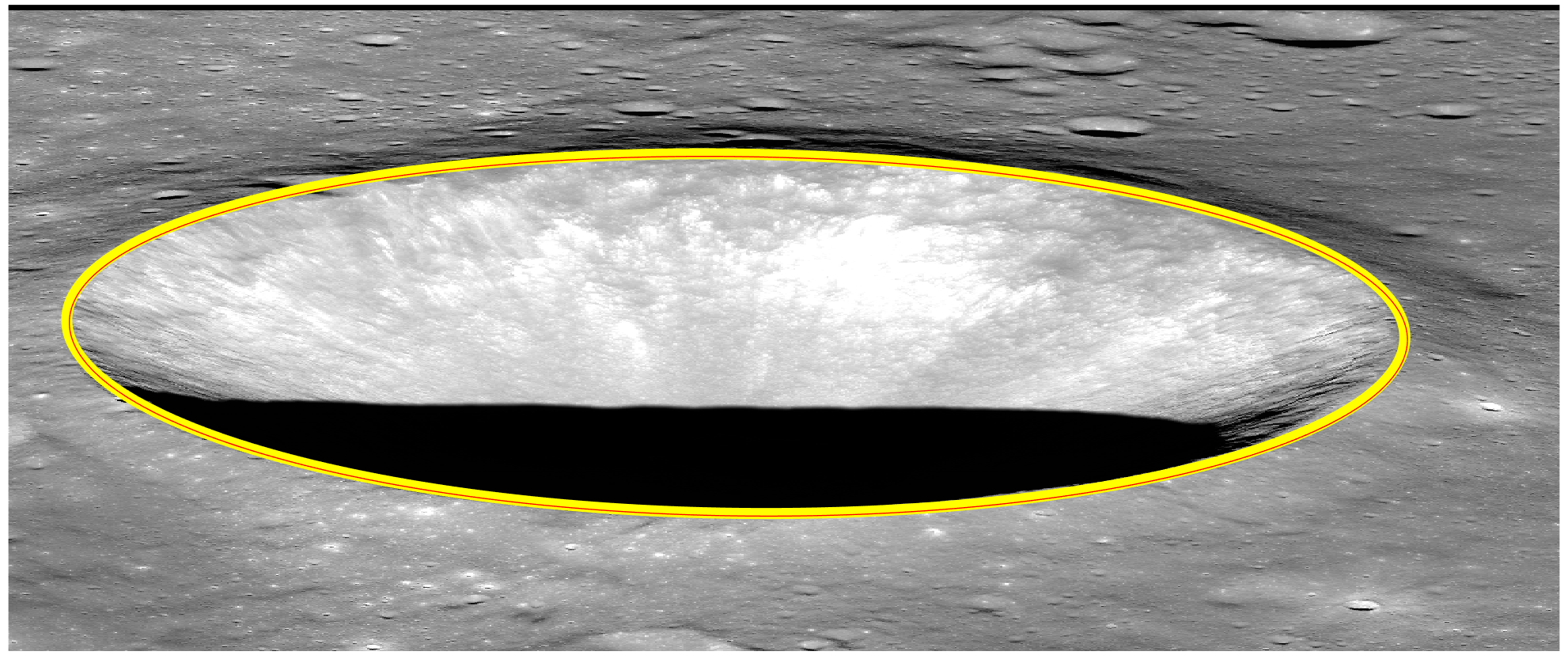}}
\caption{Zoom-in on each of the craters from Fig.~\ref{fig:all_full}. The aligned rim projection (yellow) shows an excellent qualitative agreement with the apparent crater rim shape in the image. There is a small offset in the analytic projection (red), most likely due to catalog reference frame inconsistencies and map tie errors.}
\label{fig:craters_and_curve}
\end{figure}

    Using the procedure just outlined, we determined the expected curve using Eq. \eqref{eq:ImplicitCompact}, and overlayed it on a real pushbroom image for different craters. We noticed a small offset between the estimated curve and the crater's rim position.
    This is probably due to a misalignment between the lunar frame used by PDS and the crater database. The chosen images are shown in Fig.~\ref{fig:all_full}, and a zoom on the crater and the projected curve is provided in Fig.~\ref{fig:craters_and_curve}. In this last figure, we show both the projected curve obtained from raw SPICE data, as well as the same curve after manual re-centering, which was applied to compensate for the reference frame mismatch.

For all images, we used points on the estimated (red) curve to recover the position and velocity of the spacecraft using the technique described in Sec. \ref{sec:motion_from_pushbroom_images}. The results shown in Table~\ref{tab:noiselessresults} were produced using the minimum required rim points (eight) and show errors close to machine precision.

\begin{table}[hb!]
    \centering
        \caption{Errors in the estimate of position and velocity of the spacecraft obtained using points lying on the estimated curve (yellow line in Fig.~\ref{fig:craters_and_curve})}
    \label{tab:noiselessresults}
    \begin{tabular}{c c c c c}
    \hline
          crater & Curtis (R) & Curtis (L)  & Euler L  & Hortensius\\
          \hline
          $\Delta r_{0x,C}\,[km]$ &-$4.0743\,10^{-10}$ & $3.5732\,10^{-9}$  &$1.2065\,10^{-10}$ & $ 2.5473\,10^{-11} $\\
          \(\Delta r_{0y,C}\,[km]\)& $7.5783\,10^{-10}$& $2.9511\,10^{-9}$& $2.7254\,10^{-11}$ & $ 5.6177\,10^{-14} $\\
          \(\Delta r_{0z,C}\,[km]\) &$1.5563\,10^{-7}$ & $3.6677\,210^{-7}$&$2.8745\,10^{-9}$ & $ -3.1974\,10^{-11} $\\
          \(\Delta V_x\,[km/s]\) & $2.3123\,10^{-11}$& $-1.7859\,10^{-10} $& $-1.2694\,10^{-11}$ & $ -8.7264\,10^{-14} $\\
          \(\Delta V_y\,[km/s]\) & $-4.1998\,10^{-11}$& $-1.4886\,10^{-10}$ &$-2.4084\,10^{-12}$ & $ -6.0466\,10^{-14} $\\
          \(\Delta V_z\,[km/s]\) & $-8.5903\,10^{-9}$& $-1.8560\,10^{-8} $&  $-3.0029\,10^{-10}$ & $ 1.2296\,10^{-12} $\\
          \hline
    \end{tabular}
\end{table}

\begin{figure}[h!]
    \centering
    \includegraphics[width=0.7\textwidth]{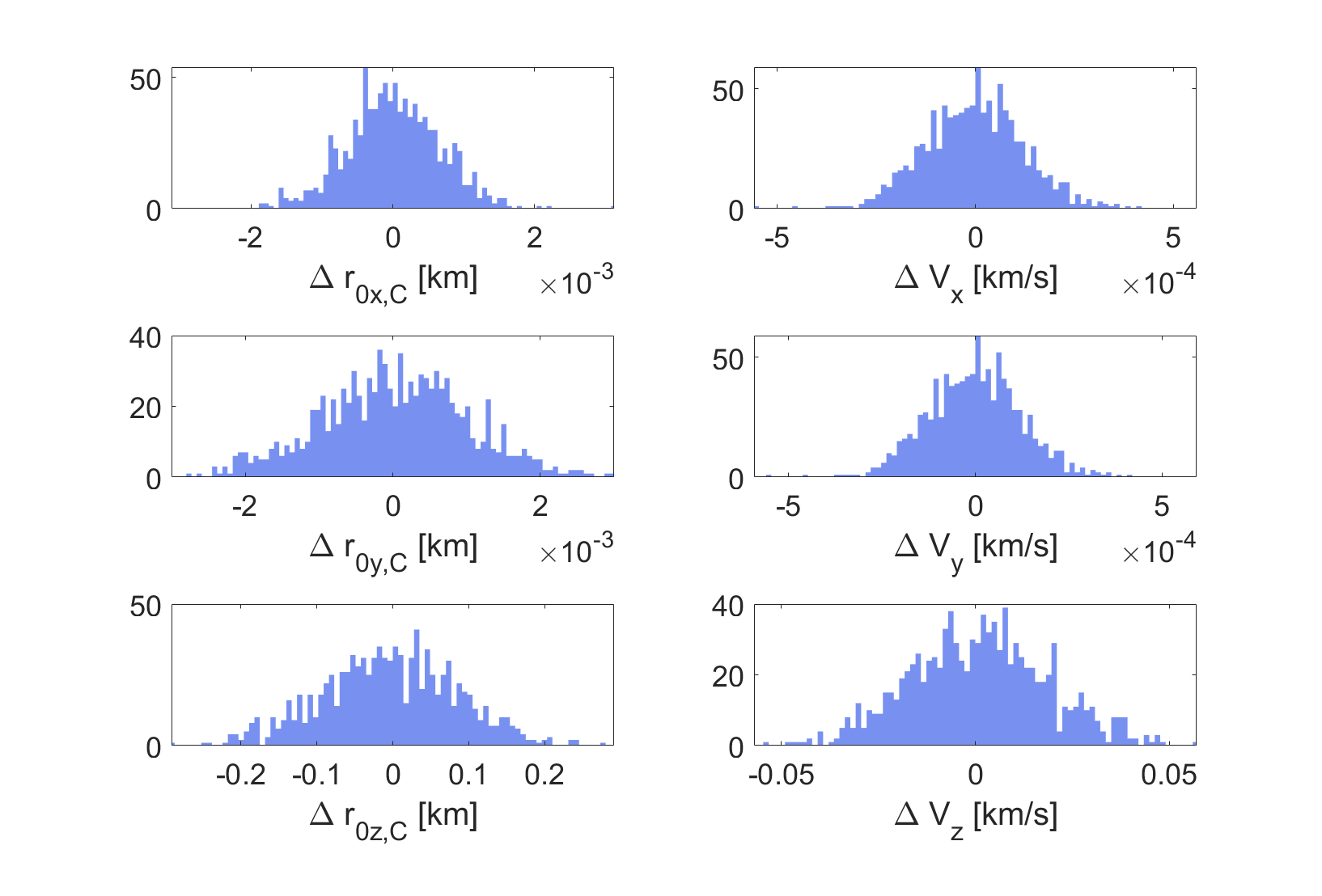}
    \caption{Results of a Monte Carlo simulation with 1000 samples, and \(\sigma=1\)~px for the Palisa A crater.}
    \label{fig:histo-Palisa}
\end{figure}

For the Palisa A crater, we also performed a Monte Carlo simulation, sampling the pixel coordinates' error from a Gaussian distribution with zero-mean and \(1\) pixel of standard deviation. The results obtained using 100 noisy points on the curve are shown in the histograms of Fig.~\ref{fig:histo-Palisa}. 

From this plot, we can see how the accuracy in the \(z\)-direction degrades more quickly than the accuracy in the \(x-y\) plane, which instead remains smaller. This indicates the sensitivity of the solver to changes in the size of the projected curve.

\subsection{Motion from real images}
In this section, we manually selected points that appear to be on the crater rim of real pushbroom images captured by the LRO NAC cameras \cite{Robinson:2010}, and we used these points to recover the state of the spacecraft. Crater rims are not always well-defined, and the user choice of which points belong to the rim has an impact on the accuracy.

\begin{figure}[h!]
    \centering
    \includegraphics[width=0.7 \textwidth] {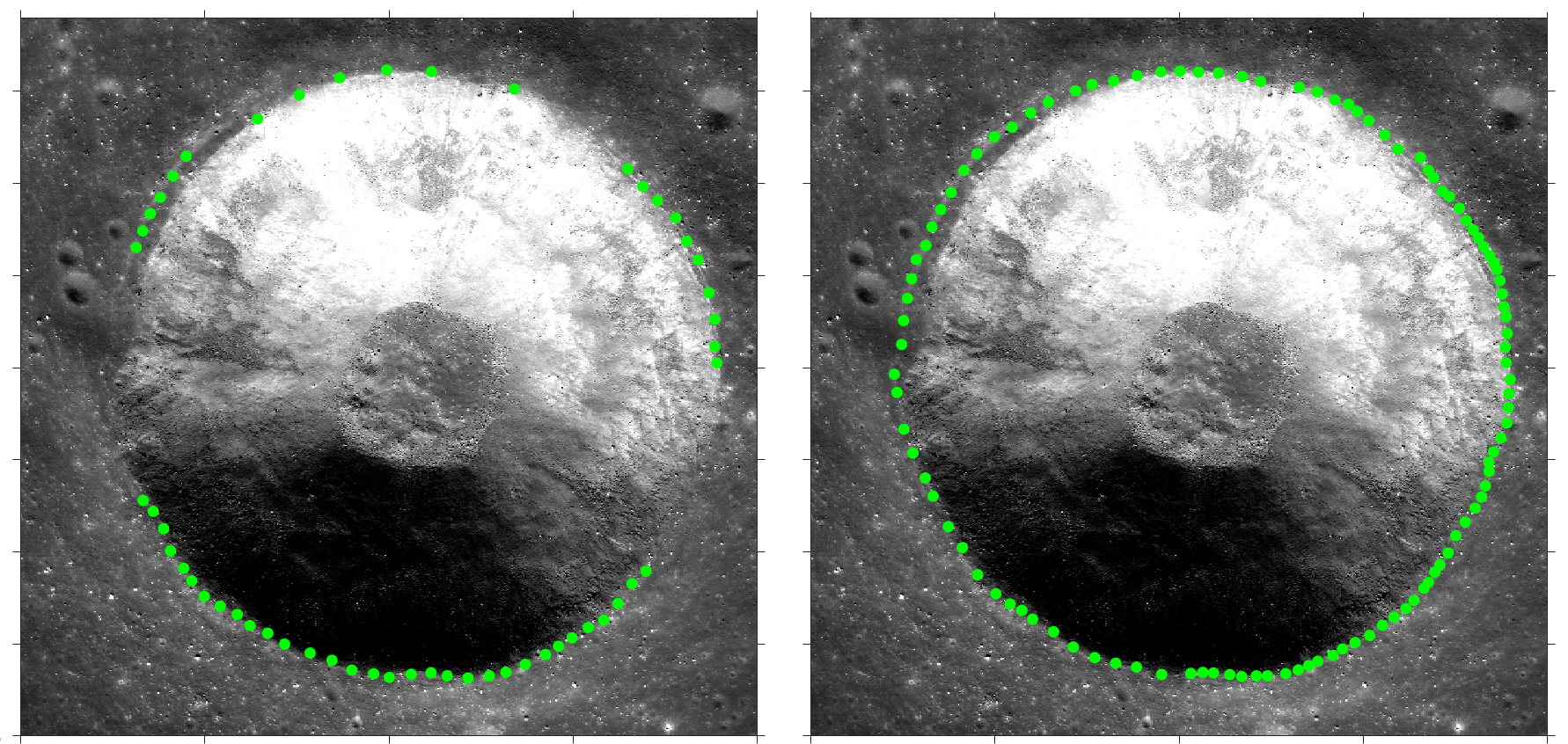}
    \caption{Selected points on the rim of the Euler L crater. On the left, 54 points. On the right, 109 points.}
    \label{fig:chosen_pointsEULER}
\end{figure}

\begin{figure}[h!]
    \centering
    \includegraphics[width=0.7 \textwidth] {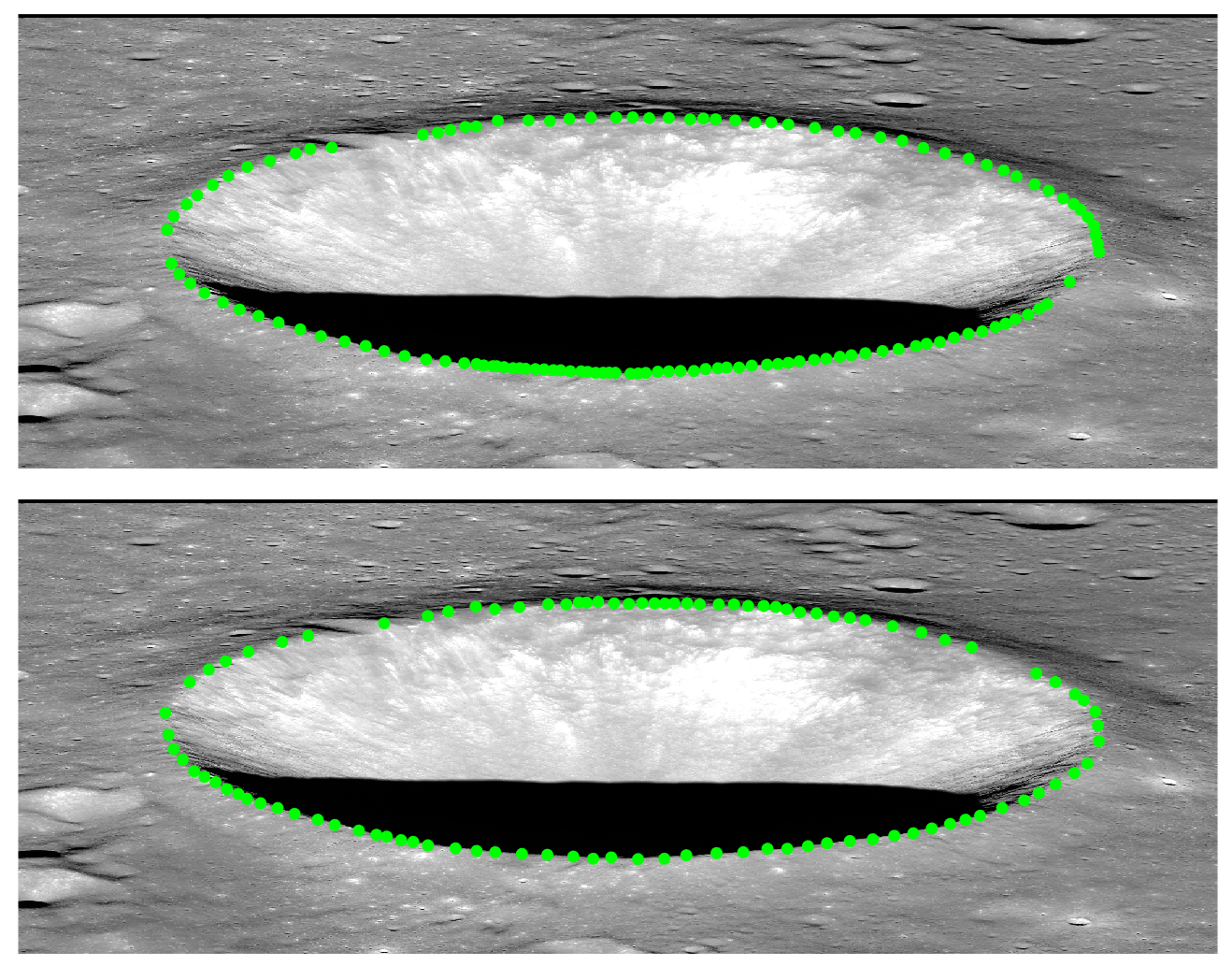}
    \caption{Selected points on the rim of the Hortensius crater. On the upper image, 125 points. On the lower image, 99 points.}
    \label{fig:chosen_pointsHORTENSIUS}
\end{figure}

Figures~\ref{fig:chosen_pointsEULER} and~\ref{fig:chosen_pointsHORTENSIUS} show two choices of points for the Euler L and Hortensius craters, respectively. The selected points produce the results shown in Table~\ref{tab:results}. As expected from the results of the Monte Carlo simulation, we notice that the estimate of the position and velocity in the \(x-y\) plane remains consistently reliable for different choices of crater rim points. However, the estimate of the state along the \(Z\)-axis of the camera frame is much more sensitive to the choice made. Notably, the respective errors indicate that a larger number of points will not always produce a better result. A way to mitigate the error in the \(z\)-direction is to select points only on portions of the rim that are clearly defined.

\begin{table}[ht!]
    \centering
        \caption{Position and velocity estimate from pushbroom images}
    \begin{tabular}{c c c c c}
    \hline
          crater &Euler L & Euler L&  Hortensius & Hortensius\\
          \hline
          points & 54 & 109 & 125 & 99 \\
          $\Delta r_{0x,C}\,[km]$  &0.1287 & 0.1894 & -0.0156 &-0.0201 \\
          \(\Delta r_{0y,C}\,[km]\)&0.3272 & 0.2542 & 0.0892 & 0.1282\\
          \(\Delta r_{0z,C}\,[km]\) &-3.0390 & 0.2955 & 1.5181& 4.5255 \\
          \(\Delta V_x\,[km/s]\)  & 0.0130 & 0.0083 & -0.0150 & -0.0146 \\
          \(\Delta V_y\,[km/s]\)& -0.0188 &-0.0112 & 0.0233& 0.0192  \\
          \(\Delta V_z\,[km/s]\) & 0.0072  &-0.2339 & -0.0352& -0.2987 \\
        \hline
    \end{tabular}
    \label{tab:results}
\end{table}

In conclusion, this approach can be used to obtain a reasonable estimate of the position and velocity of the spacecraft in the \(x-y\) plane, while the estimate of the state in the \(z\)-direction should be taken into account only if the crater rim is clearly defined.

\section{Conclusion}\label{sec:conclusion}

    In this work we develop a framework to analyze how elliptical craters project into a pushbroom camera image. We consider multiple expressions for ellipses, each of which are useful in different contexts. In addition, we relate these forms to each other via straightforward conversions. Two representations are particularly noteworthy. The first is a single-parameter explicit formulation (canonical parametric representation) that conveniently maps to pushbroom space via matrix multiplication of three basis vectors which are defined using common ellipse parameters (semi-major axis and semi-minor axis). Further, the pixel coordinates of the ellipse projection are easily obtained by varying the single canonical representation parameter, as is shown via a practical numerical example.
    
    The second noteworthy representation is the implicit form for the curve. In general, an ellipse in the world (e.g., a crater rim) projects to a polynomial of degree four in a pushbroom image. A thorough examination of the implicit form indicates that the pushbroom projection happens to be another non-degenerate conic (a polynomial of degree two) when both the 1-D pushbroom array (i.e., camera frame $Y$-axis) and the camera velocity are parallel to the surface. We tie the explicit and implicit forms in the pushbroom camera frame together with a numerical example. We do so by projecting the intersection between the quadric surface (described by the coefficients of the fourth-degree polynomial) and a saddle surface onto the plane of the pushbroom image to obtain the same result as the explicit form. Multiple experimental results show the similarities between the fourth-degree polynomial and the crater rim on real pushbroom images. 

    Finally, we develop an algorithm that can be used to recover the state of the spacecraft (position and velocity) given the knowledge of the spacecraft attitude, camera intrinsic parameters, and the pushbroom image of a known crater. We tested the algorithm by manually selecting points on the crater rim of real pushbroom images. The results show that the method produces a reliable estimate of the state of the spacecraft in the \(x-y\) plane, with an estimate of the position and velocity in the \(z\)-direction being highly dependent on the choice of points. This suggests that the estimate of the state in the \(z\)-direction should be trusted only when the crater rim is well-defined. A tool for automatic selection of crater rim points would mitigate the subjectivity related to manual selection and potentially improve the solver's reliability. 

    In summary, this paper provides an analytical framework for studying the projection of elliptical surface features (e.g., crater rims) into a pushbroom camera image, while further demonstrating how this framework can be used to recover information about the spacecraft state at the time of image capture.

\section{Acknowledgements}\label{sec:acknowledgements}
The authors thank Tara Mina for valuable feedback on this manuscript. The work of C. De Vries was partially supported through a NASA Space Technology Graduate Research Opportunities (NSTGRO) Fellowship (80NSSC22K1185). The work of J. Christian was partially supported through NASA award 80NSSC22M0151.

\section{Conflict of interest statement}
On behalf of all authors, the corresponding author states that there is no conflict of interest.
\newpage

\newpage

\appendix
    
        \section{Pushbroom Camera Model Coefficients} \label{sec:appendix}

            \begin{table}[ht!]
                \begin{center}
                    \begin{minipage}{\textwidth}
                        \caption{Coefficients for the explicit form. We are using the element-wise description of $\textbf{T}_C^M$ as given in Eq.~\eqref{eq:Telements}. Note: $\textbf{r}_{0,c}=\textbf{T}_C^M\textbf{r}_0$.}\label{tab:ExplicitCoefficients}
                        \begin{tabular*}{\textwidth}{@{\extracolsep{\fill}}cp{0.9\textwidth}@{\extracolsep{\fill}}}
                            \hline%
                            Coeff. & Expression \\
                            \hline%
                            A & $\frac{1}{\tau V_x}(aT_{11}-r_{0x,C})$ \\
                            B & $\frac{1}{\tau V_x}(2bT_{12})$ \\
                            C & $\frac{1}{\tau V_x}(-aT_{11}-r_{0x,C})$ \\
                            D & $d_y\left(-\frac{V_y}{V_x}(aT_{11}-r_{0x,C})+aT_{21}-r_{0y,C}\right)+v_p\left(-\frac{V_z}{V_x}(aT_{11}-r_{0x,C})+aT_{31}-r_{0z,C}\right)$ \\
                            E & $d_y\left(-\frac{V_y}{V_x}2bT_{12}+2bT_{22}\right)+v_p\left(-\frac{V_z}{V_x}2bT_{12}+2bT_{32}\right)$ \\
                            F & $d_y\left(-\frac{V_y}{V_x}(-aT_{11}-r_{0x,C})-aT_{21}-r_{0y,C}\right)+v_p\left(-\frac{V_z}{V_x}(-aT_{11}-r_{0x,C})-aT_{31}-r_{0z,C}\right)$ \\
                            G & $-\frac{V_z}{V_x}(aT_{11}-r_{0x,C})+aT_{31}-r_{0z,C}$ \\
                            H & $-\frac{V_z}{V_x}2bT_{12}+2bT_{32}$ \\
                            I & $\frac{V_z}{V_x}(aT_{11}+r_{0x,C})-aT_{31}-r_{0z,C}$ \\
                            \hline
                        \end{tabular*}
                    \end{minipage}
                \end{center}
            \end{table}

            \begin{table}[h!]
                \begin{center}
                    \begin{minipage}{\textwidth}
                        \caption{Coefficients for the implicit form in pixel coordinates.}\label{tab:ImplicitCoefficients}
                        \begin{tabular*}{\textwidth}{@{\extracolsep{\fill}}cp{0.9\textwidth}@{\extracolsep{\fill}}}
                            \hline%
                            Coeff. & Expression \\
                            \hline%
                            $\alpha$ & $H^2 + (G - I)^2$ \\
                            $\beta$ & $- 2EH - 2(D - F)(G - I)$ \\
                            $\gamma$ & $2(AI - CG)(G - I) - (A+C)H^2 + (G+I)BH $ \\
                            $\delta$ & $2(A+C)EH - 2(AI - CG)(D - F) - 2(AF - CD)(G - I) - B(DH + EG + EI + FH)$ \\
                            $\epsilon$ & $E^2 + (D - F)^2$ \\
                            $\zeta$ & $CAH^2 + (AI - CG)^2 + B^2GI - (AI + CG)BH$ \\
                            $\eta$ & $2(AF - CD)(D - F) - (A+C)E^2 + (D+F)BE$ \\
                            $\iota$ & $-2ACEH - 2(AF - CD)(AI - CG) - B^2(DI + FG) + B(AEI + AFH + CDH + CEG)$ \\
                            $\kappa$ & $CAE^2 + (AF - CD)^2 + B^2DF - (AF + CD)BE$\\
                            \hline
                        \end{tabular*}
                    \end{minipage}
                \end{center}
            \end{table}

            \begin{table}[h!]
                \begin{center}
                 \begin{minipage}{\textwidth}
                     \caption{Coefficients for the implicit form in image plane coordinates.}\label{tab:ImplicitCoefficients}
                        \begin{tabular*}{\textwidth}{@{\extracolsep{\fill}}cp{0.9\textwidth}@{\extracolsep{\fill}}}
                            \hline%
                            Coeff. & Expression \\
                            \hline%
                            $\hat{\alpha}$ & $4 K_1v_3'^2 - 8K_3 v_3' + 4K_6$ \\
                            $\hat{\beta}$ & $8K_3v_2' - 8K_5 + 8K_2v_3' - 8K_1v_2'v_3'$ \\
                            $\hat{\gamma}$ & $8v_1'\left(K_6r_1 - K_3r_3 - K_3r_1v_3' + K_1r_3v_3'\right)$ \\
                            $\hat{\delta}$ & $8v_1'\left(K_3r_2 - 2K_5r_1 + K_2r_3 + K_3r_1v_2' + K_2r_1v_3' - K_1r_2v_3' - K_1r_3v_2'\right)$ \\
                            $\hat{\epsilon}$ & $4K_1v_2'^2 - 8K_2v_2' + 4K_4$ \\
                            $\hat{\zeta}$ & $4v_1'^2\left(- m_2^2 + K_6r_1^2 - 2K_3r_1r_3 + K_1r_3^2\right)$ \\
                            $\hat{\eta}$ & $8v_1'\left(K_4r_1 - K_2r_2 - K_2r_1v_2' + K_1r_2v_2'\right)$ \\
                            $\hat{\iota}$ & $-8v_1'^2\left(m_2m_3 + K_5r_1^2 - K_3r_1r_2 - K_2 r_1 r_3 + K_1 r_2 r_3\right)$ \\
                            $\hat{\kappa}$ & $4 v_1'^2\left(- m_3^2 + K_4r_1^2 - 2K_2r_1r_2 + K_1r_2^2\right)$\\
                            \hline
                        \end{tabular*}
                    \end{minipage}
                \end{center}
            \end{table}

    \clearpage

\bibliographystyle{unsrt}  
\bibliography{references}  
\end{document}